\documentclass[twocolumn,showpacs,amsmath,amssymb]{revtex4}

\newcommand{\be}{\begin{equation}}  
\newcommand{\ee}{\end{equation}}

\newcommand{\bea}{\begin{eqnarray}}           
\newcommand{\eea}{\end{eqnarray}} 
\newcommand{\ba}{\begin{align}}
\newcommand{\ea}{\end{align}}
\def\de{\partial}
\def\be{\begin{equation}}
\def\ee{\end{equation}}
\def\beq{\begin{eqnarray}}
\def\eeq{\end{eqnarray}}
\newcommand{\beqn}{\begin{eqnarray*}}
\newcommand{\eeqn}{\end{eqnarray*}}
\def\lm{{\ell m}}

\def\l{{\ell }}

\def\IL{\relax{\rm I\kern-.18em L}}

\usepackage{latexsym}
\usepackage{amssymb}
\usepackage{amsfonts}
\usepackage{amsmath}
\usepackage[dvips]{graphicx}

\begin{document}

\title{Accretion--driven gravitational radiation from nonrotating compact objects: \\
 Infalling quadrupolar shells.}

\author{Alessandro Nagar,$^{1,2}$ Guillermo D\'{\i}az,$^{1}$ Jos\'e A. Pons,$^{3}$ 
        and Jos\'e A. Font$^{1}$} 

\affiliation{ $^{1}$Departament d'Astronomia i Astrof\'{\i}sica,
                     Universitat de Val\`encia, Edifici d'Investigaci\'o,
                     Dr.~Moliner 50, 46100 Burjassot (Valencia), Spain \\
              $^{2}$Dipartimento di Fisica,
                     Universit\`a di Parma,
                     Parco Area delle Scienze 7A,
                     43100 Parma, Italy \\
             $^{3}$Departament de F\'{\i}sica Aplicada,
                     Universitat d'Alacant, Ap. Correus 99, 03080
                     Alacant, Spain}

\date{\today}

\begin{abstract}
This paper reports results from numerical simulations of the
gravitational radiation emitted from non--rotating compact objects
(both neutron stars and Schwarzschild black holes) as a result of the
accretion of matter. We adopt a {\it hybrid} procedure in which we
evolve numerically, and assuming axisymmetry, the linearized equations
describing metric and fluid perturbations coupled to a fully nonlinear
hydrodynamics code that calculates the motion of the accreting
matter. The initial matter distribution, which is initially at rest,
is shaped in the form of extended quadrupolar shells of either dust or
obeying a perfect fluid equation of state.  Self--gravity of the
accreting layers of fluid is neglected, as well as radiation reaction
effects.  We use this idealized setup in order to understand the
qualitative features appearing in the energy spectrum of the
gravitational wave emission from compact stars or black holes, subject
to accretion processes involving extended objects. A comparison for
the case of point--like particles falling radially onto black holes is
also provided.  Our results show that, when the central object is a
black hole, the spectrum is far from having only one clear,
monochromatic peak at the frequency of the fundamental quasi-normal
mode. On the contrary, it shows a complex pattern, with distinctive
interference fringes produced by the interaction between the infalling
matter and the underlying perturbed spacetime, in close agreement with
results for point--like particles. Remarkably, most of the energy is
emitted at frequencies lower than that of the fundamental mode of the
black hole. Similar results are obtained for extended shells accreting
onto neutron stars, but in this case the contribution of the stellar
fundamental mode stands clearly in the energy spectrum. Our analysis
illustrates that the gravitational wave signal driven by accretion
onto compact objects is influenced more by the details and dynamics of
the process, and the external distribution of matter, than by the
quasi--normal mode structure of the central object. The gravitational
waveforms from such accretion events appear to be much more complex
than former simplified assumptions predicted.
\end{abstract}

\pacs{
04.30.Db,   
04.40.Dg,   
95.30.Lz,   
98.62.Mw    
}

\maketitle

\section{Introduction}
\label{introduction}

Most of the different astrophysical scenarios suggested as potential sources of 
gravitational radiation have in common the presence of compact objects, such as 
neutron stars, strange stars or black holes. The coalescence of a binary system 
formed by two black holes, two neutron stars or one black hole and one neutron 
star is the main target of the ground based interferometers [Laser Interferometer Gravitational
Wave Observer (LIGO), VIRGO), but 
other possibilities such as galactic supernovae are also worth exploring, especially 
in anticipation of the capabilities of future detectors. Isolated black holes, in 
particular, are reckoned to be characterized by a unique emission pattern known as 
quasi-normal mode (QNM) ringing--rapidly damped sinusoidal modes. This signal has 
been studied extensively using perturbation theory and frequency-domain techniques 
for most classes of black hole solutions (see e.g.~\cite{KokkotasSchmidt,chandra} 
and references therein). The detection of such QNM signals depends strongly on the 
luminosity of the source or, in other words, on how strongly the black hole is excited, 
but also on the knowledge of the power spectrum. It is also well known that a 
relativistic star has a very rich non-radial oscillation spectrum, and it can emit 
gravitational waves through QNM ringing (see e.g.~\cite{KokkotasSchmidt} and references 
therein). Therefore, another plausible source of gravitational radiation involves the 
rapid accretion of large clumps of matter onto a compact object (either a neutron star 
or a black hole), a recurrent and ubiquitous phenomenon in relativistic astrophysics. 
Accretion is expected to happen following the gravitational collapse of the core of 
a massive star, once a neutron star has already been formed. Part of the remaining 
stellar material, which has not been expelled by the shock driving the supernova 
explosion, may fall back onto the neutron star, until a critical mass is exceeded 
and the star collapses to a black hole. Some more material may in turn form a 
long-lived, centrifugally supported torus or disk if the collapsing star had initially 
some amount of rotation.

Semi-analytical studies of extended objects (shells or blobs of dust) falling isotropically 
onto a black hole are available in the literature~\cite{naka81b,haugan82,shapiro82,oohara83a,
oohara83b,petrich85}. These studies shed the first light in understanding the modification of 
the gravitational wave (GW) emission pattern of the black hole due to the presence of matter.
Collectively, these works showed that for a fixed amount of infalling mass $m$, the energy
released in gravitational radiation is reduced compared to the value of the point--particle 
limit ($E\sim 0.01 m^2/M$ \cite{DRPP}, $M$ being the mass of the black hole). This reduction is 
interpreted as due to cancellations of the emission from distinct parts of the extended 
object. Such conclusions were later confirmed by Papadopoulos and Font~\cite{PapadopoulosFont},
who performed numerical simulations of the gravitational radiation emitted during the 
accretion process of an extended object onto a black hole. In \cite{PapadopoulosFont} the 
first-order deviations from the exact black hole geometry were approximated by the Teukolsky 
equation~\cite{teuk72} for Schwarzschild black holes, i.e.~the inhomogeneous Bardeen-Press 
equation~\cite{bardeen}, including curvature perturbations induced by matter sources, whose 
nonlinear evolution was integrated using a hydrodynamics code. This was the first numerical 
study in the time-domain of the gravitational radiation emitted by extended objects accreted 
by black holes, and showed the gradual excitation of the black hole fundamental QNM frequency 
by sufficiently compact shells. In the thin shell limit, the energy asymptotes to a finite 
value, which is about a third of the point--particle limit.

Correspondingly, linear perturbation studies in the time domain of neutron star spacetimes, 
aimed at addressing QNM excitation, have also received considerable attention in the 
literature (see e.g.~\cite{KokkotasSchmidt} and references therein). The oscillation 
properties are obtained from the analysis of non-radial stellar perturbations. For 
non--rotating stars with polytropic equations of state, the spectrum naturally splits 
into an axial (or odd-parity) and a polar (or even-parity) part, according to the terminology 
used to address the behaviour of the perturbation equations under parity transformations. 
The axial part of the spectrum includes only gravitational modes, the so called $w$ modes, 
which are purely relativistic, being absent in Newtonian gravity \cite{KokkotasSchutz}. 
The polar part, on the other hand, contains $w$ modes as well as fluid modes. 
The excitation of axial parity modes was analyzed by Andersson and Kokkotas \cite{ak}, 
by sending pulses of gravitational waves to the neutron star and studying its response, 
and by Ferrari and Kokkotas~\cite{FerrariKokkotas}, by scattering of point-like particles. 
The excitation of polar modes was first investigated by Allen {\it et al.}~\cite{allen}, 
also as a scattering problem using gravitational wave pulses. The same approach was later 
followed by Ruoff and co-workers in \cite{RuoffI,RuoffII}, using both Gaussian pulses 
and point--particle scattering. A framework for constructing initial data sets for 
perturbations was developed in \cite{ak2} and applied to study neutron star collisions in 
the close limit approximation in Ref.~\cite{allen2}.

Alternately, in a series of papers Seidel and co-workers computed the gravitational radiation 
emanating from slightly non--spherical stellar core collapse, in the axial \cite{SeidelAxial} 
as well as in the polar case \cite{SeidelPolar}, and the waveforms associated with the formation 
of neutron stars. The zeroth-order solution was a spherical collapsing star, whose dynamics was 
computed by solving the coupled system of Einstein and hydrodynamics equations using the Lagrangian 
May-White approach. The GWs were extracted using perturbation theory on the spherical background within 
the Gerlach-Sengupta \cite{GerlachSengupta} formalism. More recently, Harada {\it et al.} \cite{harada} 
have reexamined the axial part of this problem, using null coordinates (Hernandez-Misner) and a 
gauge invariant and coordinate independent perturbative formalism developed by 
Mart\'{\i}n-Garc\'{\i}a and Gundlach \cite{Gundlach0,GundlachI,GundlachII}. Within this approach Harada 
{\it et al.} \cite{harada} have been able to follow the spherical collapse of both, supermassive stars 
and neutron stars, until a black hole forms, computing the GWs that are emitted in the process. 
Full numerical relativity computations of QNM ringing in spherical gravitational collapse using 
the characteristic formulation of general relativity are also reported in \cite{siebel}. Within 
the same formalism, the imprints of fluid accretion on the emitted GWs from a black hole were 
studied in \cite{pf2001}, finding the familiar damped-oscillatory GW decay, but both decay rate 
and frequencies being modulated by the mass accretion rate. 

In this paper, we analyze the similarities and differences in the gravitational wave emission
pattern from black holes and neutron stars as a result of the radial accretion of matter.
A detailed realistic modelization of the gravitational emission from accretion flows 
would require of three-dimensional (magneto--)hydrodynamical simulations in general relativity, 
coupled to radiation transport and diffusive processes. However, some preliminary steps can 
be taken to understand the underlying basic physics in a qualitative way before getting 
engaged in large scale computational efforts.
Our numerical procedure lies, hence, in the borderline of full numerical relativity and 
perturbation theory. As in \cite{PapadopoulosFont}, the accreting matter is evolved in a 
curved static background by solving the nonlinear hydrodynamics equations. The response of 
the compact object to the infalling matter, which triggers the emission of gravitational 
radiation, is computed using perturbation theory. More precisely, we use the gauge invariant 
formalism of \cite{GundlachI} and study the excitation of QNMs of both Schwarzschild black 
holes and neutron stars by numerically solving in the time domain the even--parity 
perturbation equations with matter sources. For the case of black holes, these equations 
reduce to the inhomogeneous Zerilli--Moncrief \cite{Zerilli,Moncrief} equation. One key 
assumption of our approach is that the mass of the accreting fluid is much smaller than 
the mass of the central compact object. Fluid {\it self--gravity} and {\it radiation 
reaction} effects are also neglected; i.e.,~we ignore the first-order metric corrections 
to the fluid equations of motion. The first approximation (no self--gravity) is in general 
valid for fluid motions in the vicinity of the compact object, where tidal forces dominate 
over the fluid self--gravity. The second approximation (no radiation reaction) is valid 
as long as the energy in the form of gravitational radiation is much smaller than the 
kinetic or internal energy of the fluid. Our procedure follows then the same {\it hybrid} 
approach previously adopted in \cite{PapadopoulosFont}, but departs from it in the formalism. 
We anticipate that, for a given numerical resolution and numerical scheme, the use of the 
Zerilli--Moncrief equation results in improved long--term numerical stability as compared 
to the Bardeen--Press equation employed by \cite{PapadopoulosFont}, allowing for an 
accurate computation of late time features of the GW signal (namely tails). The results 
reported in the present investigation are further restricted to the case of radially 
accreting shells of either dust or obeying a perfect fluid equation of state 
(EOS), where the mass density profile is shaped in the form of quadrupolar 
shells of Gaussian radial extent. In this respect this work can be considered as a necessary 
assessment of our numerical approach, in anticipation of the study of more interesting 
scenarios, namely the excitation of QNMs from perfect fluid thick accretion tori orbiting 
around compact objects, which will be presented elsewhere \cite{diazetal}.

The paper is organized as follows: Section~\ref{framework} describes in some detail the 
theoretical framework adopted, namely the construction of the unperturbed stellar models, 
the general relativistic hydrodynamics equations, the perturbation equations for neutron 
stars and black holes, and the generation of time-symmetric initial data to describe the 
accreting shells. The numerical methods used for both the hydrodynamics equations and the 
perturbation equations are outlined in Sec.~\ref{numerics}. Section~\ref{results} is 
devoted to presenting the main results of our investigation, splitting the description of 
the black hole and neutron star cases into two separate subsections. Finally, 
Sec.~\ref{conclusions} summarizes the main conclusions of this work and outlines future 
directions in this research. Appendix~\ref{appendixA} contains technical information 
regarding the general form of the source term for the inhomogeneous Zerilli--Moncrief 
equation, and Appendix~\ref{appendixB} presents a comparison with point--like particles 
falling onto black holes. We use units such that $c=G=1$.

\section{Theoretical framework}
\label{framework}

\subsection{Unperturbed stellar models}
\label{stellar_model}

The background metric of a non--rotating, spherically symmetric star is given by the 
line element
\begin{align}
ds^2 = -e^{2a}dt^2+e^{2b}dr^2+r^2\left(d\vartheta^2+ \sin^2\vartheta d\varphi^2\right),
\end{align}
where $a$ and $b$ are functions of the radial coordinate $r$ only. Assuming the star 
is a perfect fluid whose energy momentum tensor reads
\begin{align}
T_{\mu\nu}=\left(\epsilon +p \right)u_{\mu}u_{\nu}+p\,g_{\mu\nu},
\end{align}
with $p$ denoting the pressure, $\epsilon$ the total energy density, and $u^{\mu}$ the 
fluid $4$-velocity, the Einstein's equations become the Tolman-Oppenheimer-Volkoff (TOV) 
equations of hydrostatic equilibrium:
\begin{align}
\frac{d m}{d r}&=4\pi r^2\epsilon~,\\
\frac{d a}{d r}&=
\frac{\left(m+4\pi r^3 p\right)}{\left(r^2-{2mr}\right)},\\
\frac{d p}{d r}&=-\left(\epsilon+p\right) \frac{d a}{d r},
\end{align}
where $m(r)$ is the gravitational mass enclosed in a sphere of radius $r$. The surface 
of the star $r=R$ is determined by the condition $p=0$. At the exterior the geometry 
reduces to the Schwarzschild solution. The above system of ordinary differential equations 
(ODEs) can be integrated once the EOS is chosen and a value of the central energy density 
$\epsilon_c$ is specified. We further assume the fluid to be isentropic, so that we adopt a 
one--parameter EOS, $p=p(\epsilon)$, in polytropic form $p=K\epsilon^{\Gamma}$, 
where $\Gamma$ is the adiabatic exponent.

For the calculations we discuss in this work, we will consider two stellar models with 
the same gravitational mass, \hbox{$M=1.4M_{\odot}\approx 2.067$ km}, and the same 
adiabatic exponent \hbox{$\Gamma=2$}, one model being more compact than the other. 
These two models are meant to bracket the interval of possible radii of realistic 
neutron stars. The more compact model (A) has \hbox{$\epsilon_c=2.455 \times 10^{15}$
g/cm$^3$} and \hbox{$K=122.25$ km$^2$}, so that the radius is \hbox{$R=9.80$ km}. The 
less compact model (B) has \hbox{$\epsilon_c=0.92\times 10^{15}$g/cm$^3$} and 
\hbox{$K=180$ km$^2$} which leads to \hbox{$R=13.44$ km}.

Our analysis is further restricted to polar perturbations induced by external matter 
flows. For the sake of reference, the first frequencies of the polar part of the spectrum 
of the two stellar models considered are listed in Table~\ref{label:table1}. They 
have been obtained using a frequency-domain code based on the classical Lindblom-Detweiler 
formulation of the perturbed Einstein equations \cite{LindblomDetweiler} and described
in \cite{PonsStructure}. 

\begin{table}[t]
\caption{\label{label:table1}Frequencies of the first fluid and
gravitational modes for the two neutron star models considered.}
\begin{ruledtabular}
\begin{tabular}{ccccc}
&Mode      & A [kHz]  & B [kHz]& \\
\hline
&$f$       & 2.584    & 1.666 &\\
&$p_1$     & 3.948    & 4.045 &\\
&$w_1$     & 11.609   & 10.380& \\
&$w_2$     & 20.197   & 18.429& \\
\end{tabular}
\end{ruledtabular}
\end{table}
%
%
\subsection{General relativistic hydrodynamics equations}

The motion of a fluid in a curved spacetime is governed by the local conservation laws 
of baryonic number and energy momentum:
\begin{align}
\nabla_{\mu}J^{\mu}=0,\qquad\nabla_{\mu}t^{\mu\nu}=0\;,
\end{align}
where $J^{\mu}=\rho u^{\mu}$ is the mass density current and \hbox{$t^{\mu\nu}=\rho hu^\mu 
u^\nu+pg^{\mu\nu}$} is the stress energy tensor for a perfect fluid. In these expressions, 
$\rho$ is the rest-mass density, $h$ is the specific enthalpy, defined as $h=1+e+p/\rho$, 
and $e$ is the specific internal energy. The system of equations is closed with an EOS 
$p=p(\rho,e)$.

The equations of general relativistic hydrodynamics are a system of hyperbolic 
equations; as shown by~\cite{Banyuls}, it can be explicitly written as a system of
conservation laws. This is accomplished by defining quantities which are directly measured 
by Eulerian observers, i.e. the rest mass density $D=\rho W$, the momentum density 
in the $j$ direction $S_{j}=\rho hW^2v_j$ and the total energy density $E=\rho hW^2-p$. 
In these definitions, $W$ stands for the Lorentz factor, which satisfies $W=(1-v^2)^{-1/2}$, 
with $v^2=\gamma_{ij}v^iv^j$. Here, $v^i$ is the 3-velocity of the fluid, defined as 
$v^i=u^i/W+\beta^i/\alpha$, where $\alpha$ and $\beta^i$ are the spacetime lapse function 
and shift vector, respectively, and $\gamma_{ij}$ are the spatial components of the 
spacetime metric where the fluid evolves. For a generic spacetime, the system of equations 
we solve reads~\cite{Banyuls}
\begin{align}\label{FullHydro}
\dfrac{1}{\sqrt{-g}}\left(\dfrac{\de\sqrt{\gamma}{\bf U(w)}}{\de t}
+\dfrac{\de\sqrt{-g}{\bf F^{\mathit{i}}(w)}}{\de x^i}\right)={\bf S(w)},
\end{align}
where $g=\det(g_{\mu\nu})$, ${\bf U}({\bf w})=(D,S_j,E-D)$, and ${\bf w}=(\rho,v_i,e)$ is 
the vector of {\it primitive} variables. The expressions for the flux and source vectors, 
${\bf F^i(w)}$ and ${\bf S(w)}$, can be found in Ref.~\cite{Banyuls}. We note that in the 
current work we use the standard form of the Schwarzschild metric in polar-radial coordinates 
$(t,r,\vartheta,\varphi)$ to describe the exterior spacetime of the compact object. Hence, 
the above expressions are specialized accordingly.

\subsection{Stellar polar metric perturbations induced by hydrodynamical sources}
\label{StarPerturbations}
 
\subsubsection{Interior equations}
\label{InteriorEquations}

A formulation of the equations describing the polar perturbations of a star in the 
Regge-Wheeler gauge \cite{ReggeWheeler}, written in a form suitable for numerical 
simulations in the time domain, was first discussed by \cite{allen}. An alternative 
derivation of the same equations, based on the linearized Arnowitt-Deser-Misner (ADM) 
formalism, can be found in \cite{RuoffI}. Recently, building on the work of Gerlach and 
Sengupta \cite{GerlachSengupta}, Mart\'{\i}n-Garc\'{\i}a and Gundlach have developed a 
gauge-invariant and coordinate-independent formalism for non-spherical perturbations of 
spherically symmetric spacetimes 
\cite{Gundlach0,GundlachI,GundlachII}. In our work, we follow the formalism laid out 
in \cite{GundlachI} specifying the equations to the case of a static spherical star 
and choosing the Regge-Wheeler gauge \cite{ReggeWheeler}. The equations we obtain are 
then equivalent to those of Refs.~\cite{allen} and \cite{RuoffI}, although different 
metric variables are used.  Hence, for each $(\l,m)$ pair, the even-parity metric 
perturbation $\delta g_{\mu\nu}$ is parametrized by two scalar quantities, 
$k$ (the perturbed 3-conformal factor) and  $\chi$ (the actual gravitational wave 
degree of freedom), so that it reads 
\begin{align}
\delta g_{\mu\nu}=\begin{pmatrix}
(\chi+k)e^{2a} &   -\psi e^{a+b} &   0                   &   0                   \cr
-\psi e^{a+b}  & (\chi+k)e^{2b}  &   0                   &   0                   \cr
0              &          0      &   kr^2                &   0                   \cr
0              &          0      &   0                   & k r^2 \sin^2\vartheta \cr
\end{pmatrix}Y_{\lm}\;.
\end{align}
Since the background is static, $\psi$ is not an independent quantity, as it can be 
obtained by quadrature from $k$ and $\chi$~\cite{Gundlach0}; the relationship with the 
usual Regge-Wheeler variables can also be found in~\cite{Gundlach0}. We note that, although
the polar problem on a static star is known to have only two metric degrees of freedom inside 
the star and one degree of freedom outside (represented by the Zerilli-Moncrief function), 
we have decided to consider an additional variable inside the star, the perturbation 
of the relativistic enthalpy $H = \delta p/(p+\epsilon)$, as suggested by earlier studies of 
the subject~\cite{allen,RuoffI}. Correspondingly, at the exterior we evolve two (constrained) 
degrees of freedom instead of just one (see below). 
We notice, however, that successful evolution algorithms using the actual number of degrees 
of freedom have been developed in the past in more general frameworks \cite{SeidelPolar}. 
 
We formulate, then, the polar perturbation problem through a couple of hyperbolic equations 
for $\chi$ and $H$ plus an elliptic equation, the Hamiltonian constraint, which is used to 
compute $k$ at every temporal slice. This permits us to obtain the frequencies of the stellar 
pulsation modes with an accuracy comparable to frequency domain calculations.
The set of equations reads
\begin{widetext}
\begin{align}
\chi_{,tt}-e^{2(a-b)}\chi_{,rr}&=-e^{2a}\Bigg\{-2\left[2e^{2b}
\left(\dfrac{m}{r^2}+4\pi r p\right)^2+8\pi\epsilon-
\dfrac{6m}{r^3}\right](\chi+k)+\dfrac{\lambda-2}{r^2}\chi
\nonumber \\
&\qquad-\left[4\pi r(5p-\epsilon)-\dfrac{2}{r}+10\dfrac{m}{r^2}\right]\chi_{,r}\Bigg\},\\
H_{,tt}-c_s^2e^{2(a-b)}H_{,rr}&=-e^{2a}\Bigg\{
\bigg[\dfrac{m}{r^2}\left(1+c_s^2\right)+4\pi r p\left(1-2c_s^2\right)
+\left(4\pi r \epsilon-\dfrac{2}{r}\right)c_s^2\bigg]H_{,r}
\nonumber \\
&\qquad-\bigg[4\pi\left(p+\epsilon\right)\left(3c_s^2+1\right)
-c_s^2\dfrac{\lambda}{r^2}\bigg]H
\label{Hinterior}
+\dfrac{1}{2}\left(\dfrac{m}{r^2}+4\pi p r\right)\left(1-c_s^2\right)
\left(\chi_{,r}-k_{,r}\right)
\nonumber \\
&\qquad+\left[\dfrac{2\left(m+4\pi p r^3\right)^2}{r^3\left(r-2m\right)}
-4\pi\left(3p+\epsilon\right)c_s^2\right]
\left(\chi+k\right)\Bigg\},
\end{align}
\begin{align}
\label{HamiltonianInside}
e^{-2b}k_{,rr}-\left(\dfrac{\lambda}{r^2}-8\pi\epsilon\right)k
+\left(8\pi\epsilon-\dfrac{\lambda+2}{2r^2}\right)\chi-\dfrac{e^{-2b}}{r}\chi_{,r}
+\left(\dfrac{2}{r}-\dfrac{3m}{r^2}-4\pi\epsilon r\right)k_{,r}
+\dfrac{8\pi(p+\epsilon)}{c_s^2}H=0,
\end{align}
\end{widetext}
where $\lambda=\l(\l+1)$ and $c_s^2=\de p/\de\epsilon$ is the sound speed. 
At the star surface $\epsilon$, $p$ and $c_s^2$ vanish, and $m(R)=M$; 
thereby the evolution equation for $H$ reduces to the ODE
\begin{eqnarray}
H_{,tt}=-\dfrac{M(R-2M)}{R^3}\bigg[H_{,r}+\dfrac{1}{2}&\left(\chi_{,r}-k_{,r}\right)\bigg]
\nonumber\\
&-\dfrac{2M^2}{R^4}\left(\chi+k\right)\;.
\label{Hsurf}
\end{eqnarray}
Since we are setting $\Gamma=2$, the term proportional to $H$ in
the Hamiltonian constraint (\ref{HamiltonianInside}) is regular at $r=R$.

\subsubsection{Exterior equations with a general source term}
\label{ExteriorEquations}

The exterior equations in vacuum are readily obtained from the interior equations setting 
$\rho=p=0$ and $m=M$.  As we have seen, the equations for the induced metric perturbations 
are basically wave equations with potentials. The presence of an extended object outside 
the star or black hole reflects in the fact that these equations are not homogeneous 
anymore, but they contain source terms involving the stress-energy tensor of the external 
fluid, $t_{\mu\nu}$. Whereas for point-like particles these source terms can be explicitly 
computed by analytic techniques \cite{RuoffII,Zerilli}, in the case of extended objects 
they involve steps that cannot be managed analytically \cite{PapadopoulosFont}. To the best 
of our knowledge, an explicit expression of the source terms for a general stress energy
tensor has not yet been reported in the literature. 

Following the notation of Ref.~\cite{Gundlach0}, the (gauge-invariant) decomposition of 
$t_{\mu\nu}$ in polar spherical harmonics reads
\begin{align}
t_{\mu\nu}&=\sum_{\l=0}^{\infty}\sum_{m=-\l}^{\l}t^{\lm}_{\mu\nu}\\
&=\sum_{\l=0}^{\infty}\sum_{m=-\l}^{\l}
\begin{pmatrix}
T_{AB}^{\lm}Y^{\lm} &   T_A^{\lm}\left(Y^{\lm}\right)_{:a}\cr
\cr
T_A^{\lm}\left(Y^{\lm}\right)_{:a} & r^2T_3^{\lm}Y^{\lm}\gamma_{ab}+T_2^{\lm}Z_{ab}^{\lm}\cr
\end{pmatrix}\nonumber\\
\nonumber
\end{align} 
where the capital indexes run over the $M^2$ Lorentzian manifold and the lowercase indexes 
over the unit radius 2-sphere $S^2$, as the background spacetime can be written as a direct 
product $M^2\times S^2$. We also follow \cite{Gundlach0} for the definition of the scalar and 
vector spherical harmonics, $Y^{\lm}$ and $\left(Y^{\lm}\right)_{:a}$, respectively, and of 
the tensor spherical harmonic, $Z_{ab}^{\lm}\equiv Y^{\lm}_{:ab}+\dfrac{\lambda}{2}\gamma_{ab}
Y^{\lm}$. Here, the notation ${:a}$ stands for the covariant derivative with respect to the 
metric $\gamma_{ab}\equiv \mathrm{diag}(1,\,\sin^2{\vartheta})$ of $S^2$. In 
Ref.~\cite{Gundlach0}, the homogeneous equations of linearized nonspherical polar perturbations 
of a general time--dependent spherically symmetric spacetime were obtained. If we choose the 
background spacetime to be the Schwarzschild solution and interpret $t_{\mu\nu}$ as a source 
induced by a certain distribution of matter on this spacetime, one can arrive after some 
algebra at the polar perturbation equations of a Schwarzschild spacetime with source terms, 
which are given by
\begin{widetext}
\begin{align}
&\chi_{,tt}-e^{2(a-b)}\chi_{,rr}=-e^{2a}\bigg\{-\frac{2M}{r^3}\bigg[2 e^{2b}\frac{M}{r}-6\bigg]
\left(\chi+k\right)+\frac{\lambda-2}{r^2}\chi-\frac{2}{r}\left(\frac{5M}{r}-1\right)
\chi_{,r}\bigg\}+16\pi S_{\chi},\\
\nonumber\\
\label{HamiltonianConstraint}
&e^{-2b}k_{,rr}+\left(\frac{2}{r}-\frac{3M}{r^2}\right)k_{,r}-\frac{e^{-2b}}{r}\chi_{,r}
-\frac{\lambda}{r^2}\,k-\dfrac{\lambda+2}{2r^2}\chi+8\pi S_{\cal H}=0.\\
\nonumber
\end{align}
\end{widetext}
The sources $S_{\chi}$ and $S_{\cal H}$ read 
\begin{widetext}
\begin{align}
S_{\chi}:=&\;e^{2(a-b)}\Bigg\{T_{11}^{\lm}+\left(T_{2}^{\lm}\right)_{,rr}-
2\left(T_{1}^{\lm}\right)_{,r}+\dfrac{1}{r}\left(\dfrac{5M}{r}e^{2b}-3\right)
\left(T_{2}^{\lm}\right)_{,r}-\dfrac{2}{r}\left(\dfrac{3M}{r}e^{2b}-1\right)T_{1}^{\lm}\nonumber\\
&\qquad\qquad\qquad\qquad\qquad\qquad\qquad\qquad\;+\dfrac{1}{r^2}\bigg[\dfrac{2M^2}{r^2}
\,e^{4b}+8-\left(\dfrac{4M}{r}+\dfrac{\lambda+8}{2}\right)e^{2b}\bigg]T_2^{\lm}-e^{2b}T_{3}^{\lm}
\Bigg\}\;,\\
\nonumber\\
S_{\cal H}:=&\;e^{-2a}T_{00}^{\lm}\;.
\end{align}
\end{widetext}
It is known that the perturbations of the Schwarzschild spacetime are described by a 
hyperbolic equation for a single function, originally written in the frequency domain 
and in the Regge-Wheeler gauge by Zerilli \cite{Zerilli} and later in the time domain 
by Moncrief \cite{Moncrief}, but adopting a gauge invariant formulation. Following the 
normalization convention of \cite{RuoffI}, the Zerilli-Moncrief function is related to $k$ and 
$\chi$ as follows \cite{Moncrief,GundlachII}
\begin{align}\label{DefZeta}
Z = \dfrac{4 r^2 e^{-2b}}{\lambda\left[(\lambda-2)r+6M\right]}
\bigg[\chi+\left(\frac{\lambda}{2}+\frac{M}{r}\right)e^{2b}k-r~k_{,r}\bigg].
\end{align}
In presence of matter sources, this function is a solution of the inhomogeneous 
Zerilli-Moncrief equation 
\begin{align}
Z_{,tt}-e^{2(a-b)}Z_{,rr}=\dfrac{2M}{r^2}e^{2a}Z_{,r}+V_{\l} Z+S_z\;,
\label{inzerilli}
\end{align}
where the Zerilli potential $V_{\l}$ is given by
\begin{align}
&V_{\l}=-\left(1-\dfrac{2M}{r}\right)\\
\nonumber\\ 
&\times\dfrac{\lambda(\lambda-2)^2r^3
+6(\lambda-2)^2Mr^2+36(\lambda-2)M^2r+72M^3}
{r^3\left[(\lambda-2)r+6M\right]^2}\nonumber\;.
\end{align}
The source term $S_z$ appearing in Eq.~(\ref{inzerilli}) for a general matter distribution 
has not been reported in the literature in a form suitable for time domain computations. 
Its derivation can be found in Appendix A. The final result is
\begin{widetext}
\begin{align}\label{ZerilliSource}
S_{z}=-\dfrac{16\pi e^{2a}}{\lambda\left[(\lambda-2)r+6M\right]}\Bigg\{&\dfrac{ e^{-2a}}{(\lambda-2)r+6M}
\Big[\lambda\Big(6r^3-16Mr^2\Big)-r^3\lambda^2-8r^3+68Mr^2-108M^2r\Big]T_{00}^{\lm}\nonumber\\
&+e^{-2b}\Big[2Mr+r^2(\lambda-4)\Big]T_{11}^{\lm}+2r^3\left(T_{00}^{\lm}\right)_{,r}
-2r(r-2M)^2\left(T_{11}^{\lm}\right)_{,r}\nonumber\\
&+4\lambda(r-2M)T_{1}^{\lm}
+\left[2\lambda\left(1-\dfrac{3M}{r}\right)-\lambda^2\right]T_{2}^{\lm}+4r(r-2M)T_3^{\lm}\Bigg\}\;.
\end{align}
\end{widetext}
In the case of a point particle, this general source term reduces to that given in 
\cite{RuoffII}.

For any multipole $(\l,m)$ the radiated energy is computed from the Zerilli function. 
It is given by
\begin{align}
E^{\lm}&=\int_{-\infty}^{\infty}\left(\dfrac{dE}{dt}\right)^{\lm}dt\nonumber\\
&=\dfrac{1}{64\pi}\,\dfrac{(\l+2)!}{(\l-2)!}\int_{-\infty}^{\infty}\left|Z^{\lm}_{,t}\right|^2dt\;.
\label{elm}
\end{align}
By defining the Fourier transform $\tilde{Z}$ of $Z\equiv Z^{\lm}$ as
\begin{equation}\label{FourierTransform}
\tilde{Z}(\omega,r)=\int_{-\infty}^{\infty}e^{-i\omega t}Z(t,r)dt\;,
\end{equation}
we obtain the energy as follows:
\begin{align}\label{energyspectrum}
E^{\lm} &= \int_0^{\infty}\left(\dfrac{dE}{d\omega}\right)^{\lm}d\omega\nonumber\\
&=\dfrac{1}{64\pi^2}\,\dfrac{(\l+2)!}{(\l-2)!}\int_0^{\infty}\omega^2
\left|\tilde{Z}(\omega,r)\right|^2d\omega\;.
\end{align}

\subsection{Black hole polar metric perturbations induced by hydrodynamical sources}

When the neutron star is replaced by a Schwarzschild black hole, the polar 
perturbation problem becomes much simpler, since one only needs to solve the inhomogeneous 
Zerilli--Moncrief equation. As mentioned in the Introduction, in order to draw a comparison 
with the neutron star case, we consider the black hole case discussed 
in Ref.~\cite{PapadopoulosFont}, but solving Eq.~(\ref{inzerilli}) instead of the 
inhomogeneous Bardeen-Press (BP) equation. The underlying motivation behind this 
choice is the possibility of performing long-term stable evolutions that allow
for the extraction of late time features (radiative power-law tails) in the GW signals. 

Working with the same numerical method, this result seems to be unreachable with 
the BP equation, because this equation is {\em intrinsically} unstable. To make the 
argument clearer, let us recall that this equation, written using the Regge-Wheeler 
tortoise coordinate $r^*$ \cite{ReggeWheeler},
\begin{align}
r^*=r+2M\log\left(\dfrac{r}{2M}-1\right)\;,
\end{align}
reads
\begin{align}\label{BPequation}
Y_{,tt}-Y_{,r^*r^*}&-\dfrac{4(r-3M)}{r^2}\left(Y_{,t}+Y_{,r^*}\right)\nonumber\\
&=-\dfrac{\Delta}{r^4}\left[\dfrac{6M}{r}+\left(\l+2\right)\left(\l-1\right)\right]Y+\dfrac{8\pi\Delta}{r}{\cal T}\;,
\end{align}
where the (complex) function $Y$ is related to the Weyl tensor tetrad component 
by $Y=r\Psi_4$; $\Delta$ is the {\it horizon function} $\Delta=r^2-2Mr$ and 
${\cal T}$ is the source term determined by matter flows \cite{PapadopoulosFont}. 
The polar metric perturbations correspond to the real part of $Y$, the axial ones to its imaginary part.  
At $r=3M$ the term proportional to $Y_{,t}$ changes sign: depending on the relative sign 
between $Y_{,tt}$ and $Y_{,t}$, one may view $Y_{,t}$ either as a damping term 
(when the signs of both coefficients agree, i.e. for $r<3M$) or an antidamping term (otherwise). 
As argued by Krivan {\em et al.}~\cite{krivan}, who analyzed in detail the general case of 
the Teukolsky equation for Kerr black holes, this is likely the origin of 
exponentially growing modes that appear when the equation is numerically solved by standard 
finite-differencing explicit methods. Although some attempts to delay in time the onset of 
the instability have been investigated in the literature \cite{PapadopoulosFont,krivan}, 
it remains an open issue. As noted above, this effect 
is particularly disturbing in the presence of matter sources, since the instability is always 
occurring before that the late-time state of the system is reached.

On the other hand, using the tortoise coordinate, Eq.~(\ref{inzerilli}) reads
\begin{align}
Z_{,tt}-Z_{,r^*r^*}=V_{\l}Z+S_z\;,
\end{align}
where the source term is the same one given by Eq.~(\ref{ZerilliSource}), 
with the derivatives with respect to $r$ consistently replaced by derivatives 
with respect to $r^*$. This equation does not present terms which may cause
exponential growing modes, and, therefore, it permits stable evolutions.
 
\subsection{Initial data}

The nontrivial issue of how to specify suitable initial data (i.e.~gravitational 
radiation free) in the presence of sources has been addressed to some extent in a number 
of works~\cite{MartelPoisson,RuoffII,allen,allen2,PapadopoulosFont}. In particular, 
initial data suitable to describe point-like particles scattered by stars or falling 
onto black holes can be found in \cite{RuoffII} and \cite{MartelPoisson}, respectively. 
The common procedure is to choose initial data such that the Hamiltonian and momentum 
constraints are satisfied at the initial temporal hypersurface. If the matter source
is initially at rest, the initial conditions are time symmetric and the momentum 
constraint is automatically satisfied if the Hamiltonian constraint is. However, if 
velocity fields are present initially, the momentum constraint must be solved for 
too~\cite{RuoffII}.

In the simulations reported in this work, we choose time--symmetric initial configurations 
where the matter distribution, shaped in the form of quadrupolar shells of dust or
perfect fluid, is falling onto the central compact object from rest. Hence, we only 
need to consider the Hamiltonian constraint, Eqs.~(\ref{HamiltonianInside}) and 
(\ref{HamiltonianConstraint}). This constraint is a single equation for three unknowns, 
$H$, $k$, and $\chi$. By setting $H=0$, one of the functions $k$ or $\chi$ can be specified 
freely and the constraint is then solved for the remaining one. Furthermore, the initial data 
are chosen so as to minimize the amount of gravitational radiation present initially. 
This is done by choosing $\chi=0$ and solving for $k$ with a given source $T_{00}$ \cite{MartelPoisson}. 
Given $k$, the initial profile of $Z$ is then computed using Eq.~(\ref{DefZeta}). While 
this prescription should ensure that the initial data are free of any spurious GW content 
other than that due to the presence of the matter source, in practice this is not exactly 
the case, a certain amount of GWs being always present. Its origin is related to the 
finite value of the initial location of the shell. For such a configuration, the 
impossibility of specifying the GW contribution associated with the shell in a way 
consistent with its past history provokes a transient burst in which the excess GWs 
are radiated away. Similar situations were considered in 
Refs.~\cite{RuoffII,LoustoPrice,MartelPoisson}, where the source of the perturbations 
was a particle orbiting around or scattered off a star or a black hole. In order to 
minimize this problem in our simulations, we freely evolve the perturbations without 
the hydrodynamics part until the initial unphysical burst of gravitational radiation 
leaves the numerical domain, taking the final profiles as the initial state for the 
actual simulations. In the black hole case, this simple procedure permits us to avoid 
completely any kind of initial data interference. For stars, however, this approach 
triggers the oscillations of the fluid modes. Hence, we further proceed by resetting 
$H=0$, in order to obtain the correct initial model.

\section{Numerical framework}
\label{numerics}

The numerical schemes we have implemented to solve the hydrodynamics and perturbation 
equations (outlined in the preceding section) are used with some technical differences 
in both scenarios under study, neutron stars and black holes. The first difference is 
that, in the neutron star case, the overall grid is uniformly spaced in the radial 
coordinate $r$, while for the black hole it is uniformly spaced in the tortoise coordinate $r^*$. 
The numerical domain chosen to discretize the hydrodynamics equations is always smaller 
than that of the perturbation equations, which is extended in both directions, towards 
the horizon of the black hole (or the origin of coordinates in the case of the star) 
and towards large radii. This procedure avoids or minimizes the effect of the spurious reflection of 
waves at both boundaries. For stars, the hydrodynamics domain begins at the first cell 
outside the stellar surface, and extends up to $r_{\mathrm{max}}=108$ km. We choose the 
same resolution ($\Delta r\sim 0.07$ km) for both stellar models, so that the interior 
is covered with 146 points for model A and 200 points for model B which, we recall, is 
less compact. The hydrodynamics grid is covered with 1400 zones, a resolution chosen to 
ensure convergence. The external wave domain extends up to $r\sim 1500$ km and is covered 
by roughly 22000 zones. On the other hand, in the case of black holes the hydrodynamics 
domain starts very close to the horizon so as to include as much as possible the peak of 
the potential barrier as well as its falloff toward $r^*=-\infty$. The sensitivity of our 
numerical results to the location of the inner boundary of the hydrodynamics domain is 
discussed in Sec.~\ref{results} below.

Before turning to describe the numerical schemes implemented in the code, it is worth 
commenting that in a realistic scenario the mass of the compact object would grow in 
time as the accretion process proceeds. Such a possibility, however, has not been 
considered in the present simulations. Technically, this would require recomputing at 
each evolution time step the equilibrium structure of the star with a mass $M+\delta 
M(t)$ (or, analogously, increasing the black hole mass). In our simulations we simply 
assume that $\delta M\ll M$, neglecting the effect of a dynamically growing mass. 

\subsection{Evolution of the external fluid}

The hydrodynamics part of our code is the same as that used in the simulations reported
in Ref.~\cite{PapadopoulosFont}. In this code in order to evolve dynamically the infalling 
fluid shells, the general relativistic hydrodynamics equations are solved using a 
Godunov-type scheme (see e.g.~\cite{Banyuls} for details). In axisymmetry 
($\partial_{\varphi}=0$), the vector ${\bf U}$ of evolved quantities appearing in
Eq.~(\ref{FullHydro}) is updated from time level $t^n$ to $t^{n+1}$ according to a 
conservative algorithm
\begin{align}\label{conservedAlgorithm}
{\bf U}^{n+1}_{i,j}&={\bf U}^n_{i,j}-\dfrac{\Delta t}{\Delta r}
\left[\hat{\bf F}^r_{i+1/2,j}-\hat{\bf F}^r_{i-1/2,j}\right]\nonumber\\
&-\dfrac{\Delta t}{\Delta\vartheta}
\left[\hat{\bf F}^{\vartheta}_{i+1/2,j}-\hat{\bf F}^{\vartheta}_{i-1/2,j}\right]+
\Delta t{\bf S}_{i,j}\;,
\end{align}
where $\Delta t=t^{n+1}-t^n$, and $\Delta r$ and $\Delta\vartheta$ indicate the radial 
and angular grid spacing, respectively. In practice, a conservative, second-order, 
two-step Runge-Kutta algorithm is employed instead of Eq.~(\ref{conservedAlgorithm}). 
In the above equation $i$ and $j$ label the radial and angular zones, respectively. 
The numerical fluxes (e.g. ${\bf\hat{F}}^r_{i+1/2,j}$) are calculated at every cell interface 
using an approximate (linearized) Riemann solver built upon the characteristic information 
of the Jacobian matrices of the system. The reader is addressed to \cite{Banyuls} for 
further details. 

The matter model we choose for the accreting shells can be either dust or perfect fluid.
In the former case, as $p=0$, the Riemann solver must be specialized accordingly to avoid
divergences. For shells accreting onto black holes we employ dust, not only to allow for
a direct comparison with the results of~\cite{PapadopoulosFont}, but also because of  
numerical difficulties encountered when using a perfect fluid in regions very
close to the event horizon. In Schwarzschild coordinates some 
metric components blow up at the horizon, which affects the evolution of hydrodynamical 
quantities. In particular, the coordinate flow velocity becomes ultra-relativistic and 
reaches the speed of light at the horizon, making the Lorentz factor infinite. If the inner 
boundary needs to be placed very close to the horizon to capture the fall off of the Zerilli 
potential (we have checked that $r^*=-50M$ is a reasonable value; see below) the metric 
and hydrodynamical quantities, despite being regular, present steep radial gradients, which 
make the numerical evolution difficult. In our current code and with the maximum grid 
resolutions affordable, we have found severe limitations in the perfect fluid accretion case 
to move the innermost boundary past values of about $r^*\sim -3M$ which, as we show below 
for the case of dust shells, are not yet close enough to the horizon to capture the 
gravitational wave emission unambiguously. We note that this numerical drawback can be 
entirely removed by using {\it horizon-adapted} coordinate systems such as ingoing 
Eddington-Finkelstein coordinates employed in simulations of perfect fluid accretion 
onto black holes in Ref.~\cite{hacs}. Such a procedure for perfect fluid accretion may be 
attempted in future work.

However, for the case of neutron stars we do not encounter numerical difficulties in evolving
perfect fluid shells. Therefore, in those simulations we do not consider quadrupolar dust shells, 
but only perfect fluid shells. This is also further motivated from the reflecting boundary conditions 
we impose at the surface of the star (see next paragraph), which may naturally lead to the appearance 
of shock waves which could not be treated in the case of dust.

The boundary conditions to impose upon the hydrodynamics variables completely depend on the 
case under study. In the black hole case, we adopt ingoing radial boundary conditions at the 
innermost grid point of the hydrodynamics domain, chosen as close as possible to the event 
horizon. As for the neutron star, the boundary conditions to impose to the infalling fluid 
shell as it arrives at the stellar surface should take into account the interaction between 
the shell and the star. As a result of the complexity in modeling this phenomenon within our 
perturbative approach, we impose reflecting boundary conditions 
at the inner edge of the hydrodynamics domain (i.e.~the surface of the star), so that 
the star is seen by the external fluid as a {\it hard} sphere. This choice includes the most 
relevant effect; that is, the pressure gradient stops the infalling matter and avoids the violation 
of energy conservation that outgoing boundary conditions would introduce (as happens in the 
black hole case when the inner boundary is not close enough to the event horizon).
Correspondingly, at the outermost radial boundary we impose stationary values of the
so-called Michel solution at those grid points~\cite{michel} at all times. We note that
the stationary Michel solution is used in the entire hydrodynamics grid in order to provide 
a dynamically unimportant, spherically symmetric accreting {\it atmosphere} surrounding the 
fluid shell. In the angular direction, axial symmetry fixes the appropriate boundary 
conditions at the axis ($\vartheta=0$ and $\pi$).

\subsection{Integration of the perturbation equations}
\label{NumericsPerturbations}

\subsubsection{Black hole case} 

We can solve the Zerilli--Moncrief equation using either a standard three-level
leapfrog method, at second-order in both space and time, or a second-order Lax-Wendroff 
scheme. Both methods have proved robust and stable enough for long time evolutions. 
In particular, they allow the computation of long-term features in the GW signal, 
namely the distinctive power-law tails following the black hole ringdown. 
However, when the leapfrog method is used, we find some high-frequency numerical noise 
of small amplitude at the very end of the tail. This numerical noise is not present 
when using the Lax-Wendroff scheme. Therefore, all results reported below for black hole 
spacetimes are obtained using this latter method. We note, however, that the second-order 
leapfrog produces noise-free results in the neutron star case. The particular form of such 
a scheme is discussed in the following section.

In order to apply the Lax-Wendroff method, the Zerilli--Moncrief equation is written as a 
first order hyperbolic system as follows: 
\begin{align}
\partial_{t}{\bf U}+\partial_{r^*}{\bf F}={\bf S},
\end{align}
with
\begin{align}
{\bf U}=\begin{pmatrix}
Z     \cr
w     \cr
\end{pmatrix},
\quad
{\bf F}=\begin{pmatrix}
Z      \cr
-w     \cr
\end{pmatrix},
\quad
{\bf S}=\begin{pmatrix}
w           \cr
V_{\l}Z+S_z \cr
\end{pmatrix},
\end{align}
where we have introduced the variable $w=Z_{,t}+Z_{,r}$. The time 
update algorithm is given by 
\begin{align}
{\bf U}^{n+1}_{j}&={\bf U}^n_j-\dfrac{\Delta t}{2\Delta r^*}
\left[{\bf F}^{n}_{j+1}-{\bf F}^n_{j-1}\right]\nonumber\\
&+\dfrac{\Delta t^2}{2\Delta r_*^2}\bigg[{\bf F}^n_{j+1}-
2{\bf F}^n_j+{\bf F}^n_{j-1}\bigg]\mathbb{A}+\Delta t {\bf S}^n_{j}\;,
\end{align}
where the matrix $\mathbb{A}$ reads
\begin{align}
\mathbb{A}=\begin{pmatrix}
1  &  0 \cr
0  &  -1 \cr
\end{pmatrix}\;.
\end{align}
To ensure stability of the code, the time step $\Delta t $ must satisfy the usual 
Courant-Fredrichs-Levy (CFL) condition, i.e. \hbox{$\Delta t = \lambda_{\mathrm{CFL}} 
\Delta r^*$}, with $\lambda_{\mathrm{CFL}}<1$. 

The same radial resolution is used for the two domains (hydrodynamics and perturbations) 
present in the computational grid, namely $\Delta r^*\sim 0.04M$. The angular domain of 
the hydrodynamics grid extends from 0 to $\pi$ and it is covered with 20 zones. As 
mentioned before, the hydrodynamics domain is within the wave domain, which is much larger. 
It extends from $r^*=-50M$ to $r^*=30M$ and is covered with about 2000 zones. The peak of 
the Zerilli potential is located at $r\sim 3.1M$ ($r^*\sim 1.9M$). Since the potential 
decays exponentially towards the horizon, $r^*=-50M$ is a sufficiently small value to 
minimize the effects of the truncation on the GW signal (see below). On the other hand, 
in the positive $r^*$ direction it is not necessary to extend the hydrodynamics grid much 
beyond the position of the center of the shell, because the shell is always collapsing 
towards the black hole. Correspondingly, the grid for the perturbation equations extends 
from $r^*\sim -876M$ to $r^*\sim 1250M$. Standard Sommerfeld outgoing-wave conditions are 
imposed at the external boundary, and ingoing-wave conditions at the black hole horizon. 
Obtaining an optimal resolution in the simulations requires us to use $2 \times 10^4$ points 
from $r^*=0$ towards the horizon and some $3 \times 10^4$ zones in the opposite direction. 
Such a large number of zones, however, does not imply any numerical limitation in the code 
as solving the Zerilli--Moncrief equation is a one--dimensional problem.

\subsubsection{Neutron star case}

The numerical algorithm used for solving the stellar perturbation equations does not use the 
gauge-invariant metric perturbation variable $\chi$, but instead $S=\chi/r$ (the equations 
are rewritten accordingly). The reason for this modification is that $\chi$ grows 
proportionally to $r$ for $r\rightarrow\infty$, while the amplitude of $S$ remains finite. 
This is an important property that helps to avoid unphysical instabilities of a numerical 
nature. The evolution equations for the independent variables $S$ and $H$ are discretized 
on a uniformly spaced grid and solved using the three-level leapfrog method. The remaining 
elliptic equation for $k$ is also discretized in space and then written as a tridiagonal 
linear system, which is solved at each time step inverting the corresponding matrix by
a standard $LU$ decomposition. 

\begin{figure*}
\begin{center}
\includegraphics[width=80 mm, height=72 mm]{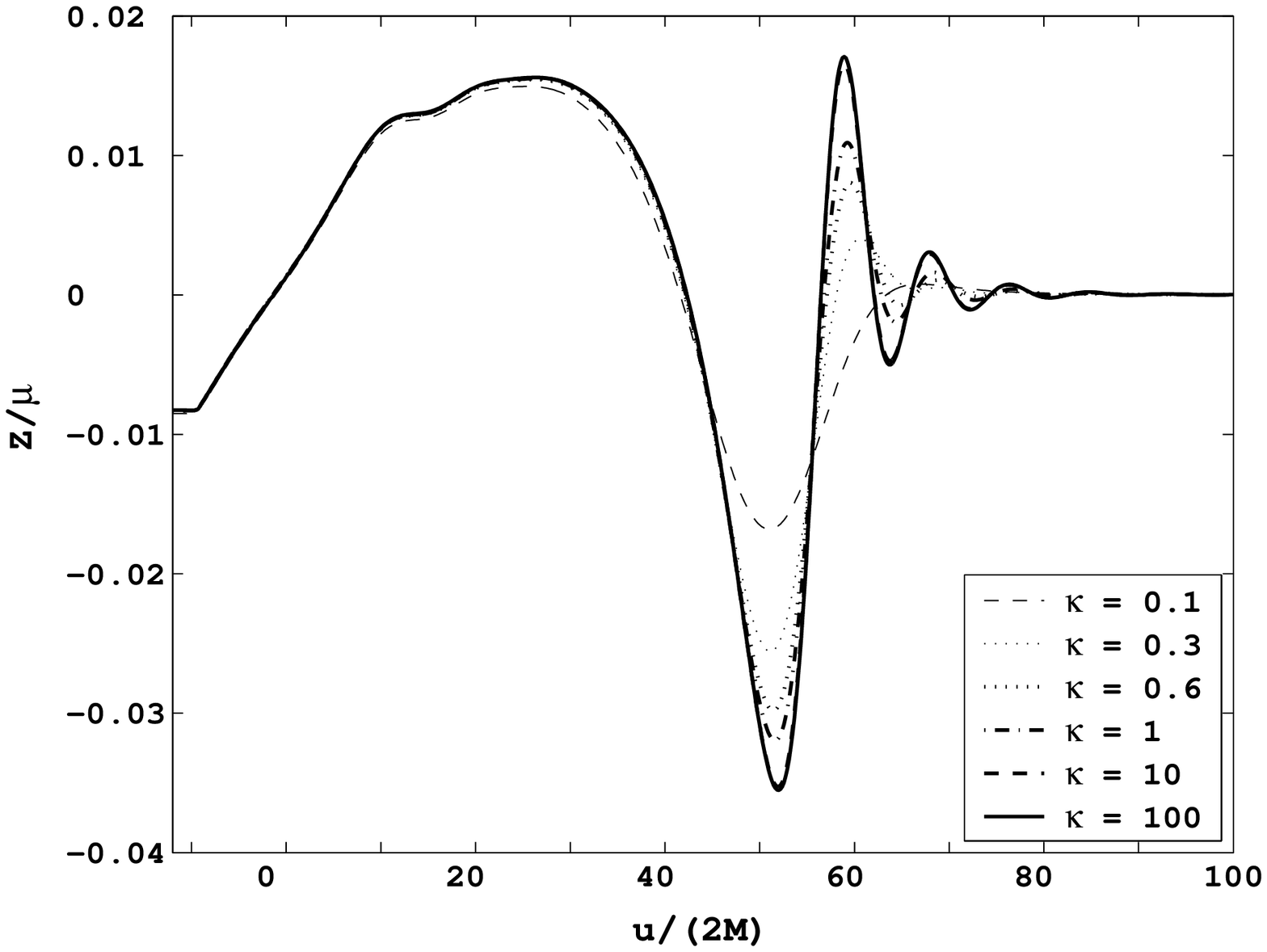}\qquad
\includegraphics[width=80 mm, height=72 mm]{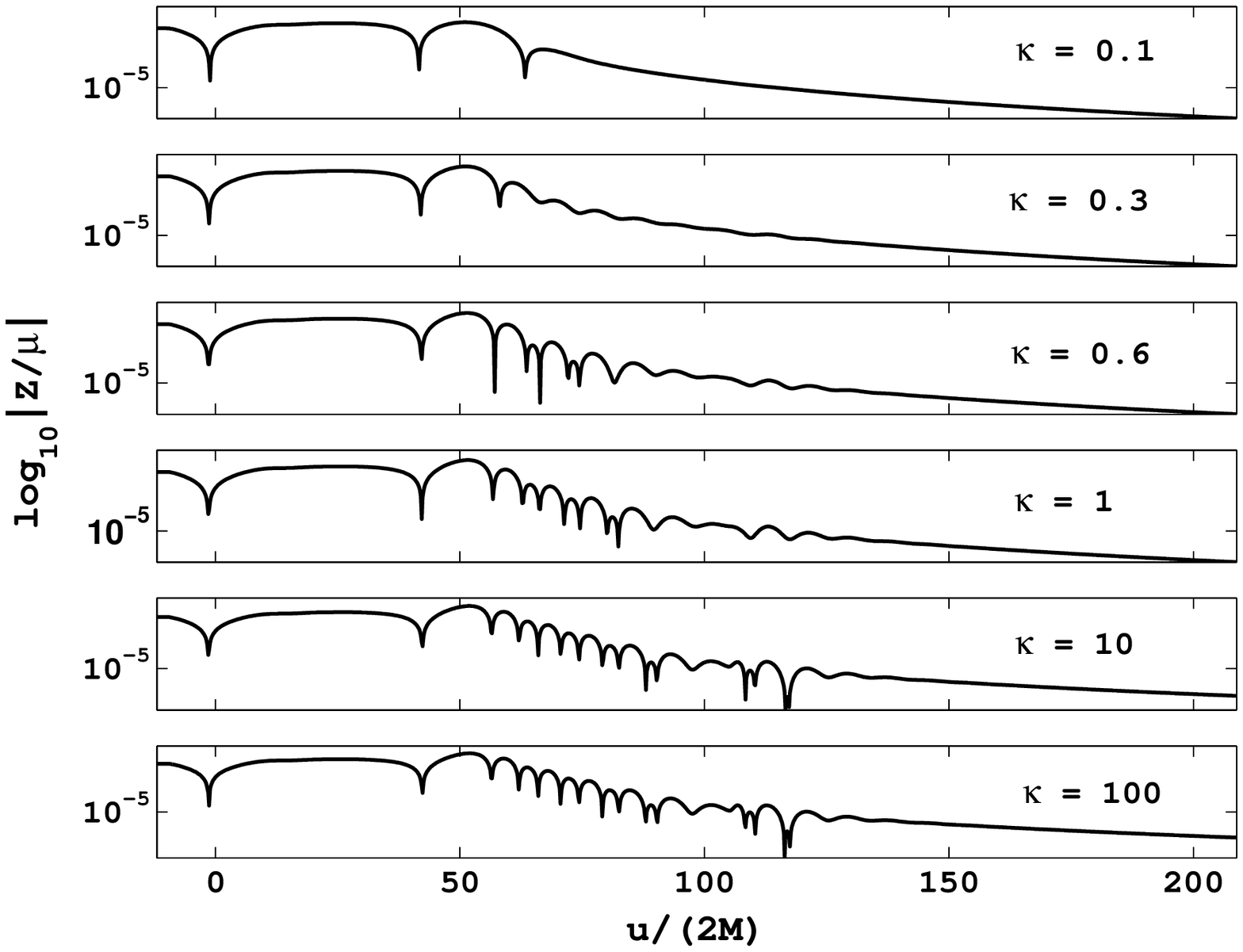}
\caption{\label{label:fig1} Gravitational waves emitted by a Schwarzschild black hole excited 
by infalling quadrupolar shells of given compactness $\kappa$ [see Eq.~(\ref{shell})]. The 
right panels show the logarithm of the GW signals shown in the left panel in order to highlight 
the onset of the fundamental QNM ringing. The shells are falling from $r_0=15M$ and the observer 
is located at $r_{\mathrm{obs}}=125M$ from the origin of coordinates.} 
\end{center}
\end{figure*} 
In order to avoid numerical problems at the origin, the left interface of the first grid 
zone is chosen to coincide with the center of the star, $r=0$, while the surface of the star, 
$r=R$, is located at a cell center. The first point of the grid, $j=1$, is then located at 
$r=\Delta r/2$, where $\Delta r$ is the radial grid spacing. Correspondingly, the surface, 
which is labeled by the $J$ cell, is located at $R=\Delta r/2+(J-1)\Delta r$. As 
mentioned before, $S$ is evolved both inside and outside the star, while $H$ is only 
evolved inside the star according to Eq.~(\ref{Hinterior}) and at the surface with 
Eq.~(\ref{Hsurf}).  The same grid spacing is chosen inside and outside the star.

The radial derivatives of $S$, $k$, and $H$ are discretized to second order,
\begin{align}
\left[A_{,rr}\right]^n_j &= \left(A^n_{j+1}-2A^n_{j}+A^n_{j-1}\right)/\Delta r^2,\\
\left[A_{,r}\right]^n_j  &= \left(A^n_{j+1}-A^n_{j-1}\right)/2\Delta r
\end{align}
(with $A\equiv S, k$ or $H$), but at $j=J$ we use backward second-order differencing for 
$H_{,r}$ as
\begin{align}
\left[H_{,r}\right]^n_{J}=\left(H_{J-2}-4H_{J-1}+3H_{J}\right)/2\Delta r\;.
\end{align}
Similarly, the second time derivatives of $S$ (and $H$) are approximated as
\begin{align}
\left[S_{,tt}\right]^n_j&=\left(S^{n+1}_j-2S^n_j+S^{n-1}_j\right)/\Delta t^2.
\end{align}

As for the black hole case, we impose standard Sommerfeld outgoing-wave boundary conditions 
at the outermost radial zone. Partial reflection from the external boundary is still present, 
but it can be minimized (or causally disconnect its influence) by placing the outermost point 
sufficiently far. At the origin of the radial coordinate $r=0$, all fields are regular and 
vanish~\cite{GundlachI}. For $\l=2$, as the origin is chosen to lie at the interface of the 
first cell, there is no need for regularizing the fields, as had to be done in~\cite{allen,RuoffI}. 

In order to have an internal consistency test, we have studied the evolution of a generic 
enthalpy profile in the source-free case, computing the Zerilli--Moncrief function in two 
different ways. In the first option, it is computed algebraically at every temporal slice from 
$k$ and $\chi$ by Eq.~(\ref{DefZeta}), all over the external domain; in the second possibility, 
it is matched to $k$ and $\chi$ using Eq.~(\ref{DefZeta}) just at the stellar surface, and then 
it is evolved independently in the exterior region by solving Eq.~(\ref{inzerilli}). Both 
solutions agree very well, and only minor differences are observed during transient states.
 
\section{Results}
\label{results}

\subsection{Black hole simulations}
\label{ResultsBH}

As mentioned in the Introduction, time-dependent numerical simulations of the accretion 
of dust shells falling isotropically onto a Schwarzschild black hole were presented 
in~\cite{PapadopoulosFont}. The aim of those simulations was to characterize and estimate 
the gravitational radiation emitted in accretion events of finite--size objects and to 
compare the outcome with the point particle case amenable to semi--analytic investigations.

The time domain simulations discussed in \cite{PapadopoulosFont} were based on the 
Newman-Penrose formalism, from which a master wave equation for black hole perturbations 
was derived by Teukolsky~\cite{teuk72}. As mentioned above, in the case of a Schwarzschild 
black hole, the Teukolsky equation reduces to the BP equation.  Here, we start by reexamining 
the GW emission of a Schwarzschild black hole as a result of the radial accretion of a 
quadrupolar dust shell. In this work, however, we use the inhomogeneous Zerilli--Moncrief 
equation for two different reasons: firstly, because it provides a direct comparison with the 
neutron star case and, at the same time, allows for cross--checking our results with those of 
\cite{PapadopoulosFont}; secondly, the main motivation in reexamining this problem was the 
fact that the Zerilli--Moncrief equation allows for long and stable time evolutions. 
This property permits the investigation of long time features in the GW spectrum. 

We consider an external matter distribution of total mass $\mu$, much smaller than that 
of the black hole, namely $\mu = 0.01M$. As in \cite{PapadopoulosFont}, the shell density 
distribution is parametrized according to
\begin{align}\label{shell}
\rho = \rho_0+\rho_{\mathrm{max}}e^{-\kappa(r-r_0)^2}\sin^2\vartheta\;,
\end{align}
where $r_0$ is the initial position of the center of the shell and $\kappa$ controls its 
width.
The background density $\rho_0$ is chosen to be very small ($\sim 10^{-22}$ km $^{-2}$) 
to simulate the vacuum outside the black hole. The angular structure of the matter density 
is assumed to have a quadrupolar profile, being given by~$\sin^2\vartheta$. The value 
of $\rho_{\mathrm{max}}$ is obtained from the condition that the volume integral of 
Eq.~(\ref{shell}) equals $\mu$. Papadopoulos and Font~\cite{PapadopoulosFont} analyzed 
the dependence of the black hole QNMs excitation on the various parameters of the shell, 
namely its width, initial location, and its initial velocity field. Here, we begin by 
studying the excitation of the black hole QNMs varying the compactness of the shell, i.e. 
its width $\kappa$ and the position where it is at rest. We restrict ourselves first to shells 
which are falling from a fixed location $r_0=15M$  and are initially at rest. The distant 
observer is located at $r_{\mathrm{obs}}=125M$ from the origin of coordinates ($r^*\sim 
133.5 M$).

The response of the black hole as a result of the accretion process when shells of 
different widths $\kappa$ are considered is shown in the left panel of 
Fig.~\ref{label:fig1}, in which we plot the time evolution of the Zerilli 
function. Note that in this figure we use retarded (observer) time $u=t-r^*$. The right 
panel shows the absolute values of the waveforms, but in logarithmic scale and separating 
the different models for clarity. The values of $\kappa$ are chosen in order to cover 
the range in which qualitative differences among the signals are observed. The process 
can be divided into three phases: (i) motion of the shell before the bulk reaches the 
peak of the Zerilli potential, from $u/2M\sim -10$ up to $u/2M\sim 50$, where the emission 
is purely due to the variation of the quadrupolar moment of the shell; (ii) shortly after $u/2M\sim 50$, 
when the bulk of the shell approaches the peak of the potential, $Z$ experiments a rapid 
variation due to the interaction with the potential barrier, resulting in a burst--like short signal. 
For more compact shells, higher amplitudes of the Zerilli function are obtained. (iii) The 
last part of the signal, from $u/2M \gtrsim 55-60$ onward, is characterized by highly 
damped oscillations at early times (the fundamental QNM ringing of the black hole), 
followed by the power-law radiative tail at the end. In the right panel of Fig.~\ref{label:fig1} 
it is visible how the onset of the fundamental QNM ringing occurs only when $\kappa \ge 0.3$; 
i.e.,~wider (less compact) shells do not succeed in exciting the fundamental mode of the central black 
hole, but rather the signals show a large wavelength oscillation. 
This feature is in excellent agreement with the results reported by~\cite{PapadopoulosFont} 
using the BP equation. If $\kappa$ is too small, the impinging GWs cannot be fully transmitted 
beyond the potential barrier to cause the ringing of the fundamental QNM. Thus, the large 
wavelength oscillation is determined by the gravitational pulse driven by the shell, which is 
almost completely reflected back by the potential barrier. It is worth mentioning that, also 
in this case, the black hole spacetime reacts to the external perturbation, as we found that 
the other QNMs, characterized by frequencies lower than the fundamental one \cite{Leaver}, 
can be slightly excited during the process. However, as a result of their high damping times, the energy 
that can be released through GWs is negligible with respect to that generated by the variation of 
the quadrupolar moment of the shell during the infall (roughly three orders of magnitude lower 
for $\kappa=0.1$). We note that the interaction of the shell with the black hole potential, 
even for small values of $\kappa$, is confirmed by the presence of the distinctive late-time decay. 

\begin{figure}[t]
\begin{center}
\includegraphics[width=80 mm, height=72 mm]{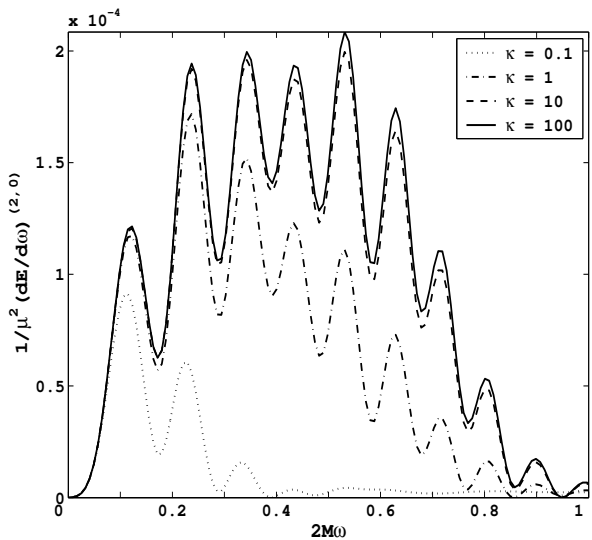}
\caption{\label{label:fig2}Energy distribution for some selected values of $\kappa$ of 
the waveforms shown in Fig.~\ref{label:fig1}. The power spectrum has a complex shape 
characterized by various distinctive peaks resulting from the interference between emitted 
and backscattered GWs (see text for details).}
\end{center}
\end{figure}

\begin{figure}[t]
\begin{center}
\includegraphics[width=80 mm, height=72 mm]{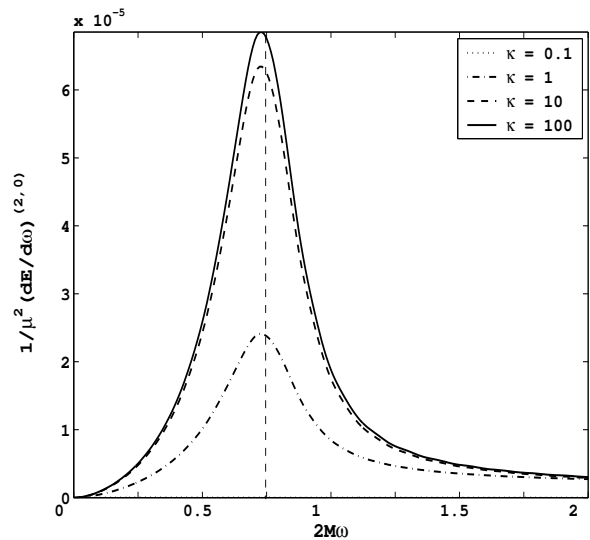}
\caption{\label{label:fig3}Excitation of the $\l=2$ fundamental QNM as a function of $\kappa$.
The spectrum is computed selecting only the part of the signal corresponding to the ringing. 
The dashed vertical line indicates the fundamental QNM frequency ($2M\omega \approx 0.7473$)
of a Schwarzschild black hole~\cite{Leaver}. Note that for $\kappa=0.1$ the spectrum is
barely visible at the bottom of the plot.}
\end{center}
\end{figure}
It is known that the generation of the fundamental QNM ringing is associated with the peak 
of the potential~\cite{CPMI}. The bulk of the shell crosses the peak at time $t/2M\sim 50$, 
which is at retarded time $u/2M\sim 49$. As argued before, the maximum GW amplitude and the 
following QNM ringing are reached somewhere around this point. The end of the accretion process 
occurs at time $t/2M\sim 79.5$, when the center of the bulk $r_{0}$ reaches the innermost boundary 
of the hydrodynamical numerical domain. This time corresponds to a retarded time $u/2M\sim 104.5$. 

The energy spectra for some selected values of $\kappa$ are shown in 
Fig.~\ref{label:fig2}. For all values of $\kappa$, the spectrum displays a complex 
structure, with several distinctive peaks that are more or less evenly spaced. 
The spacing between the maxima is roughly given by $0.1$ (in units of $2M\omega$). 
The presence of these peaks is interpreted as an effect of the interference between the 
gravitational waves emitted by the shell during its motion and the radiation emitted at 
earlier times, which has already been backscattered by the potential. This interference 
fringes are not new. In fact, similar patterns were found by Lousto and Price \cite{LoustoPrice} when 
studying the signal emitted by a point particle falling radially onto a Schwarzschild 
black hole from a finite distance $r_0$. However, some differences between their case 
and ours must be stressed. In Ref.~\cite{LoustoPrice}, the initial data had some GWs 
content, and the authors argued that the evenly spaced bumps found in the energy spectra 
(whose amplitude and spacing depended on $r_0$) were mainly due to the interference 
between the initial data pulse and the GWs emitted by the infalling point particle during 
its motion. In Appendix~\ref{appendixB}, we confirm this result using a time domain code, 
by comparing the GWs energy spectra generated by a falling point particle with and without 
the initial GW content. In the latter case, we find that the amplitude of the interference 
bumps is reduced. For the case of an imploding shell, despite the initial data effect being
eliminated by construction, we find well--defined interference fringes in the energy spectra. 
The extended size of the matter distribution results in the existence of interference patterns 
even when the GW initial content is minimized.

As for the case of point particles \cite{LoustoPrice}, we notice that the part of the 
waveforms which strongly contributes to these features in the spectrum is that extending 
up to the second zero of the signal ($u/2M\sim 42$), just before the burst. This portion 
of the waveform carries the imprint of the radiation emitted during the accretion process, 
before the bulk of the shell crosses the peak of the Zerilli potential. This is confirmed 
by the results shown in the spectra of Fig.~\ref{label:fig3}, obtained considering 
only that part of the signal from the third zero of $Z$ onwards. This helps eliminate 
as much as possible the GWs contribution related to the motion of the shell and to dig 
out the actual QNM ringing signal. The spectrum of this late time signal closely 
corresponds to that of a Schwarzschild black hole radiating via the fundamental mode 
($2M\omega\approx 0.7473$), whose frequency is marked by the vertical dashed line in
the figure. The width of the peak is consistent with the damping time of the fundamental 
mode (see~\cite{Leaver}), confirming that, if the black hole is successfully excited, 
the energy is radiated mostly in the fundamental mode.
%
\begin{figure*}[t]
\begin{center}
\includegraphics[width=85 mm, height=72 mm]{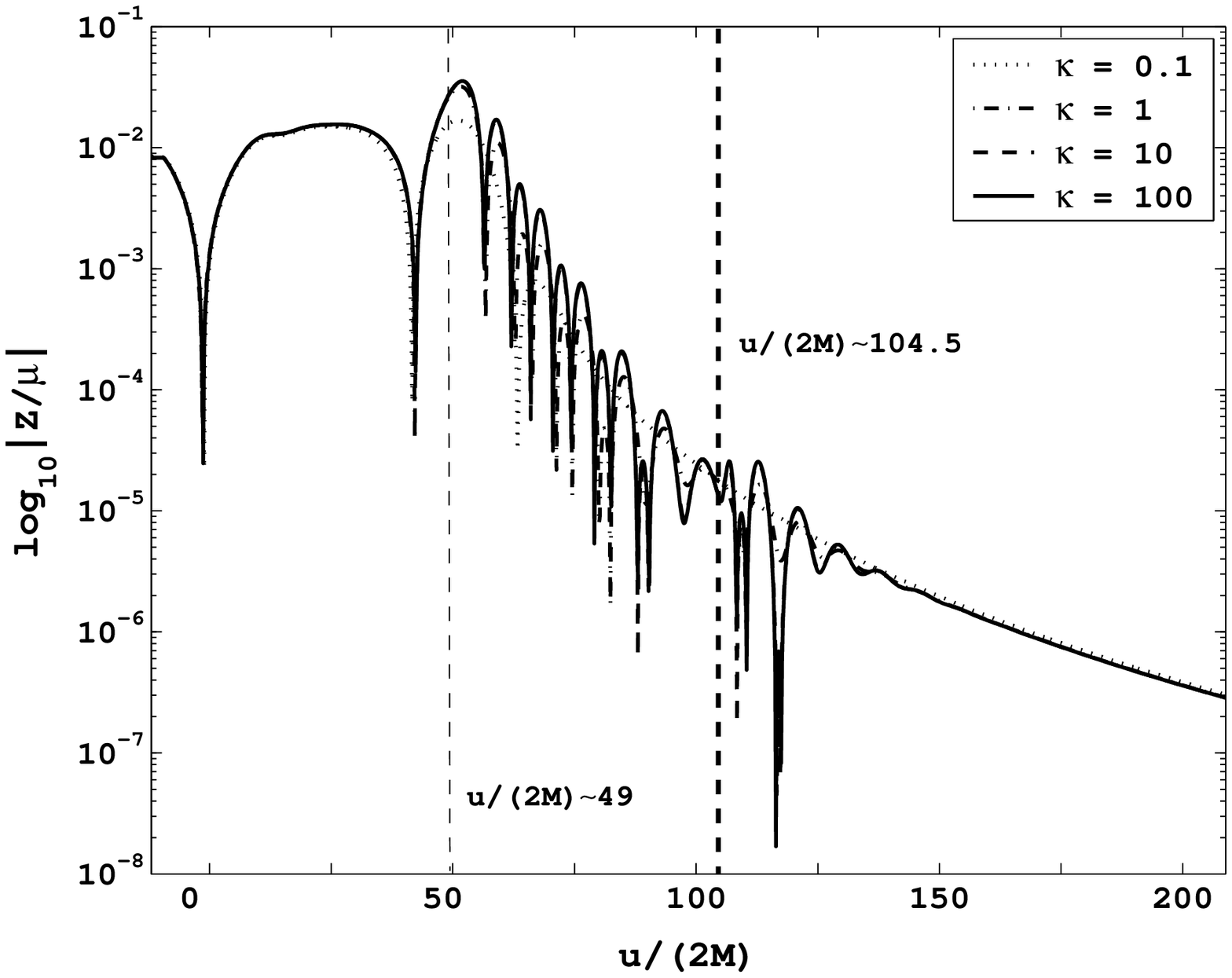}\qquad
\includegraphics[width=85 mm, height=72 mm]{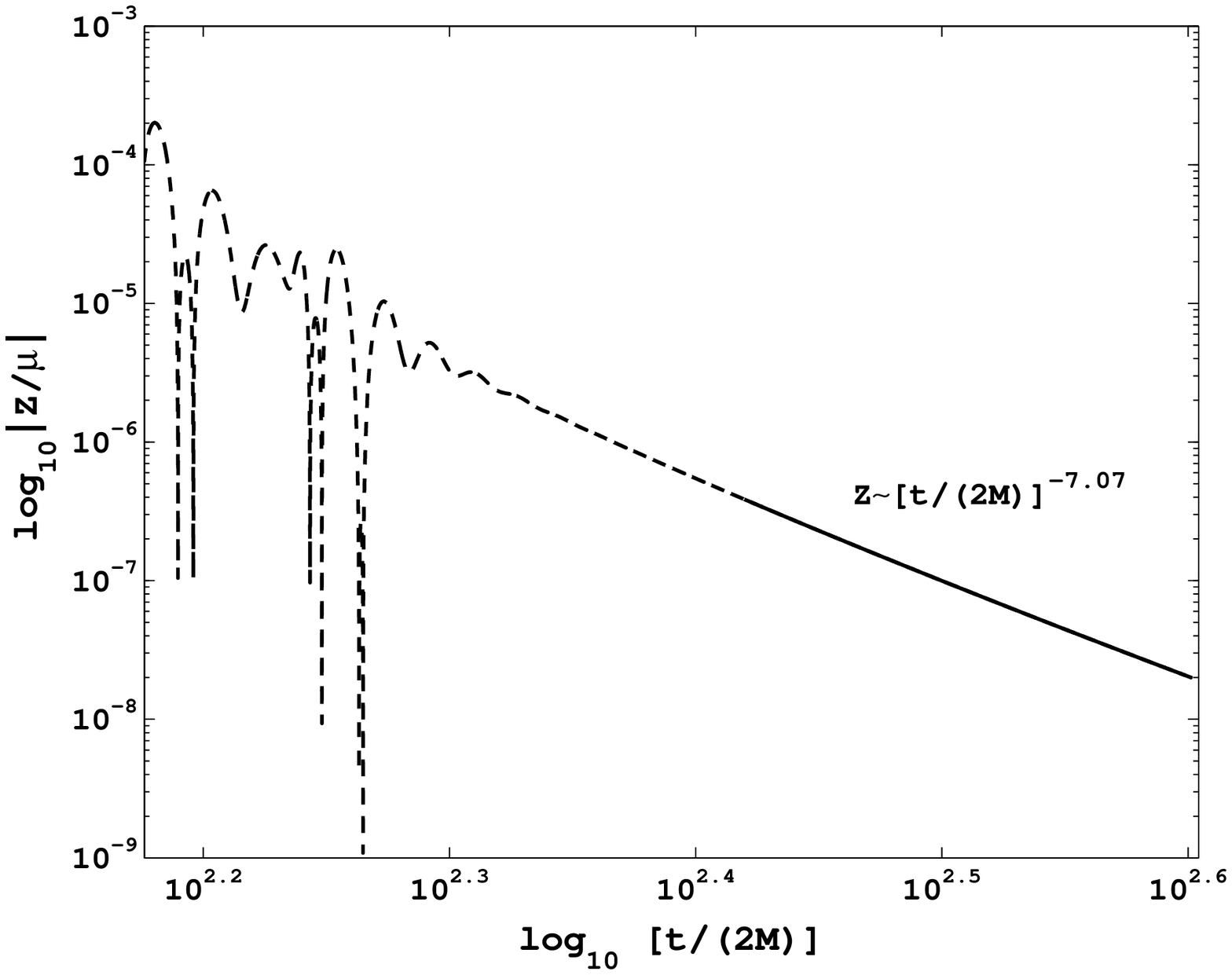}
\caption{\label{label:fig4}
Black hole QNMs ringdown and tails for various shell widths $\kappa$. The left panel shows
that the GW burst resulting from the bulk of the shell crossing the peak of the Zerilli
potential occurs at retarded time $u/2M\sim 49.5$ (first vertical dashed line). The
late-time power-law tails are all perfectly superposed for the various values of
$\kappa$ considered. At $u/2M\sim 104.5$ (second vertical dashed line), when the shell
leaves the hydrodynamical numerical domain, a second ringing appears. This ringing is a
numerical artifact (see text for details). The right panel depicts on a log-log plot the
late time behavior of the longest simulation for a shell with $\kappa=10$. A fit to the
solid line gives  $Z\sim (t/2M)^{7.07}$, in excellent agreement with the analytic
late--time falloff derived by Price \cite{Price}, Eq.~(\ref{PriceLaw}).}
\end{center}
\end{figure*}

\begin{figure}[t]
\begin{center}
\includegraphics[width=80 mm, height=72mm]{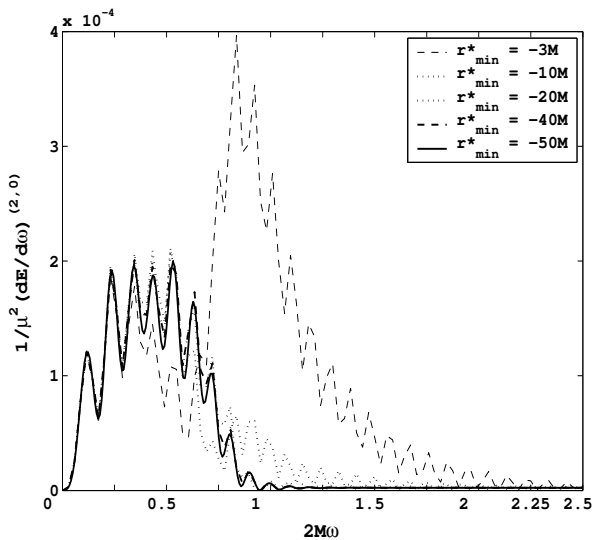}
\caption{\label{label:fig5}
Dependence of the GW energy spectra on the radial location of the innermost boundary 
$r^*_{\mathrm{min}}$ of the hydrodynamics numerical domain. The spectra plotted correspond 
to $\kappa=10$ quadrupolar shells falling from $r_0=15M$, the GWs being extracted at $r_{\mathrm{obs}}
=125M$. The unphysical high-frequency components disappear as $r^*_{\mathrm{min}}$ becomes 
smaller than about $-20M$. The spectra converge for values of $r^*_{\mathrm{min}}$ as small 
as $-50M$ (solid line).}
\end{center}
\end{figure}

In Fig.~\ref{label:fig4}, we show the superposition of the QNM ringdowns for
some selected values of $\kappa$.  As expected, the power law of the late time 
tail does not depend on the shell width $\kappa$, being all nicely overlapped. 
This important feature  was not accessible to the simulations reported 
in \cite{PapadopoulosFont}, due to the appearance of numerical instabilities
when solving the BP equation for sufficiently long evolution times. The late time behavior 
of gravitational perturbations was first studied in detail by Price~\cite{Price} using 
analytic techniques. The most recent and exhaustive discussion of this topic can be found 
in~\cite{Ching}. The gravitational multipole perturbations with $\l\ge 2$ are expected 
to fall off at large $t$ as 
\begin{equation}\label{PriceLaw}
Z\sim \left(\dfrac{t}{2M}\right)^{-(2\l+3)}\;,
\end{equation}
i.e.~as $\sim t^{-7}$ in our case ($\l=2, m=0$). The best fit to the tails shown 
in the left panel of Fig.~\ref{label:fig4}, which correspond to simulations that extend up 
to $\sim 3.67$ ms of evolution ($u/2M \sim 209$), gives $Z\sim (t/2M)^{-7.56}$. The somewhat large 
difference with the expected analytic value given by Eq.~(\ref{PriceLaw}) is due to the 
fact that the signal has not reached yet the late-time state. In order to prove this, 
we perform a much longer run (up to $u/2M \sim 609$) employing a larger grid of 
$7\times 10^4$ zones, which corresponds to roughly twice the number of zones used in the 
previous simulations. For this run, we set $\kappa=10$, keeping the same values for the 
remaining parameters. The right panel of Fig.~\ref{label:fig4} shows the results of this 
long run. The best fit to the late time signal is now $Z\sim (t/2M)^{-7.07}$, in close 
agreement with the expected value. In order to further improve the numerical results it 
simply suffices to perform even longer simulations employing larger grids.

Some more comments are relevant about Fig.~\ref{label:fig4}. For $\kappa=1$, 10, 
and 100 the signals present a strange feature, with a second ringing starting at $u/2M\sim 
104.5$ (indicated by a thick vertical dashed line in the left plot). 
We have checked that this second ringing starts at the time when the center of the shell 
leaves the numerical domain through the innermost boundary. The appearance of this ringing seems 
to be a numerical artifact, as a small discontinuity in the fields is introduced when the center 
of the shell leaves the grid. The black hole reacts to this by emitting GWs in the form of the second 
ringing, until the late time tail is reached. This unphysical ringing could be avoided by 
extending the numerical domain as much as possible towards $r^*=-\infty$ (i.e.~towards the 
event horizon). In practice, since the accretion processes we are interested in happen outside 
the horizon, where the peak of the Zerilli potential stands, it suffices to choose the innermost 
boundary of the hydrodynamics domain such that the exponential falloff of the potential 
is properly captured. The observation that most of the energy is released at low frequencies 
shows that the choice of the radial extent of the hydrodynamics domain with respect to the 
width of the Zerilli potential is of paramount importance to obtain the correct GW signals and 
the corresponding energy spectra. This effect is more important in the case of an extended 
shell, whose size changes with time due to the presence of tidal forces which tend to 
disrupt it before being swallowed by the black hole. It is the complex interaction with 
all the structure of the potential which determines the GW emission.

In order to study how the radial extent of the hydrodynamical domain affects the waveforms 
and energy emission, we focus on an accreting shell of fixed width $\kappa=10$ and vary the
radial location of the innermost boundary $r^*_{\mathrm{min}}$. The results of these 
simulations are shown in Fig.~\ref{label:fig5}. We start with $r^*_{\mathrm{min}}=-3M$ 
and gradually push $r^*_{\mathrm{min}}$ towards the event horizon, choosing the values $-10M$, 
$-20M$, $-40M$, and $-50M$. Although the low--frequency part of the spectra ($2M\omega < 0.4$) 
is almost unaffected by the location of the inner boundary, high--frequency components become 
evident when $r^*_{\mathrm{min}}$ is larger than $-20M$. The energy spectra for 
$r^*_{\mathrm{min}}=-40M$ and $r^*_{\mathrm{min}}=-50M$ are in practice identical. This means 
that such a radial extent suffices to capture all the relevant GW physics of the accretion 
process. However, using a less conservative value for the radial location of the inner boundary,
e.g.~$r^*_{\mathrm{min}}=-3M$, results in an artificial spectrum which is roughly peaked again
around the frequency of the fundamental mode. This was observed in~\cite{PapadopoulosFont}
and attributed to QNM ringing. However, it is of completely numerical nature, with the same 
origin as the second ringing discussed in Fig.~\ref{label:fig4}: when the shell crosses the 
inner boundary and leaves the grid, matter is artificially removed from the system, which 
violates energy conservation and produces the excitation of the black hole normal modes. If 
this happens at a relatively large radius such as $r^*=-3M$, the effect is amplified. Moving 
the inner boundary closer to the event horizon shifts the second QNM ringing to later times, 
when the signal is much weaker, and highly reduces the unphysical high--frequency part of the 
power spectrum. Therefore, the energy spectrum of the GW emission that one could expect in a 
realistic astrophysical scenario, during anisotropic accretion of matter onto a Schwarzschild 
black hole, is most likely to be a collection of interference fringes covering a wide range of 
low frequencies, than a single peak at the frequency of the fundamental mode of the black hole.

\begin{figure}[t]
\begin{center}
\includegraphics[width=80 mm, height=72mm]{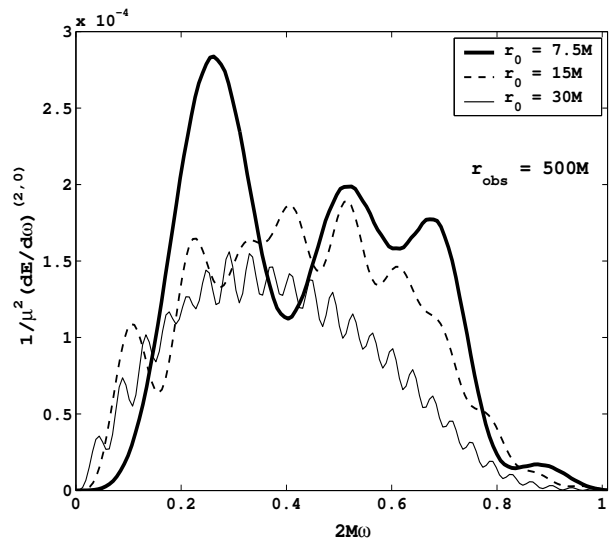}
\caption{\label{label:fig6}Dependence of the energy spectra on the initial location of
the shell ($\kappa=10$), $r_0=7.5M$ (thick solid line), $r_0=15M$ (dashed line), and $r_0=30M$ 
(thin solid line). The number of interference fringes rapidly increases with the initial 
location of the shell.}
\end{center}
\end{figure}
%
Next, we analyze the dependence of the energy spectra on the initial location of the 
shell. This is shown in Fig.~\ref{label:fig6} where we compare the spectra for three different
initial locations, $r_0=7.5M, 15M$, and $30M$. The number of interference fringes rapidly 
increases with distance $r_0$. Furthermore, the correlation between the separation of the 
fringes and the initial position of the shell is evident: the larger the distance, the smaller 
the separation between consecutive maxima. In order to explain these modulations in the spectrum, 
we can follow the reasoning of Lousto and Price~\cite{LoustoPrice}. Given a GW signal 
$Z_0(t,r^*)$, its Fourier transform $\tilde{Z}_0(\omega,r^*)$ is defined according to 
Eq.~(\ref{FourierTransform}). For large $r^*$, where the observer is located, the waveform 
represents only outgoing radiation, i.e.~$Z_0\equiv Z_0(t-r^*)$, which gives
\begin{align}
\tilde{Z}_0(\omega,r)=\int_{-\infty}^{\infty}e^{-i\omega t}Z_0(t-r^*)dt=e^{-i\omega r^*}A(\omega)\;,
\end{align}
where
\begin{align}
A(\omega)=\int_{-\infty}^{\infty}e^{-i\omega u}Z_0(u)du\;.
\end{align}
Let us now consider another GW signal $Z_1$ whose time delay with $Z_0$ is 
$T_{\mathrm{shift}}$. This signal is $Z_1\equiv Z(r^*,t-T_{\mathrm{shift}})$,
so that its Fourier transform reads
\begin{align}
\tilde{Z}_1(\omega,r^*)&=\int_{-\infty}^{\infty}e^{-i\omega t}Z(t-T_{\mathrm{shift}},r^*)dt\nonumber\\
&=e^{-i\omega T_{\mathrm{shift}}}e^{-i\omega r^*}A(\omega)\;. \nonumber
\end{align}
Thus, the Fourier transform of the signal given by the superposition of $Z_0$ and $Z_1$ is 
given by
\begin{align}
\tilde{Z}(\omega,r^*)&=\tilde{Z}_0(\omega,r^*)+\tilde{Z}_1(\omega,r^*)\nonumber\\
&=e^{-i\omega r^*}A(\omega)
\left(1+e^{-i\omega T_{\mathrm{shift}}}\right)\;.
\end{align}
Then, when computing the spectrum we have
\begin{eqnarray}
\omega^2\left|\tilde{Z}(\omega,r^*)\right|^2
=4\omega^2|A(\omega)|^2 \cos^2\left(\dfrac{1}{2} \omega T_{\mathrm{shift}}\right).
\end{eqnarray}
Therefore, the modulation in the frequency spectrum is related to the characteristic time 
$T_{\mathrm{shift}}$ which accounts for the delay between consecutive GW signals, $\delta\omega=
2\pi/T_{\mathrm{shift}}$. From the measure of the peak spacing $\delta\omega$, we can thus infer 
the corresponding time shift. In the case of a radially infalling point particle, Lousto and 
Price~\cite{LoustoPrice} argued that this time shift roughly coincides with the infalling time.
They use this empirical criterion as a rule of thumb to predict the variation of the spectra 
as $r_0$ is increased.  We have found a similar correlation for our extended dust shells. 

To close this section, we compute the total energy emitted in gravitational waves for shells
accreting from the three initial locations considered previously, \hbox{$r_0=7.5M$}, $15M$, 
and $30M$. The energy emitted is computed by integrating in frequency the energy spectra of 
Fig.~\ref{label:fig6}, where the integrals are calculated using a standard trapezoidal rule. 
As we do in the point--particle case (see Appendix~\ref{appendixB}), it is convenient to use as 
a reference quantity the ratio $2M/\mu^2 E^{20}$. Table~\ref{label:table2} lists the values of 
the energy for the three positions $r_0$ considered. We note that the third value (for $r_0=30M$) 
may be slightly underestimated due to some inaccuracies in the resolution of the power spectrum. 
It is worth stressing that, irrespective of the location $r_0$ of the shell, the values reported 
in Table~\ref{label:table2} are smaller by roughly two orders of magnitude than those obtained 
in the point--particle limit~\cite{LoustoPrice,MartelPoisson} [Lousto and Price~\cite{LoustoPrice} 
give $(2M/\mu^2)E^{20}=1.64\times 10^{-2}$ for $r_0=30M$ and $1.43\times 10^{-2}$ for $r_0=10M$]. 
The reduction we find in the total energy emitted in gravitational waves is a consequence of 
the finite size of the shells considered in the present work. 

We note that in the numerical simulations reported in~\cite{PapadopoulosFont} the estimation 
of the energy yielded considerably larger values than the ones reported here, asymptoting 
towards a third of the point--particle limit~\cite{DRPP} as the compactness of the shell was 
increased. We argue that such a high value is overestimated, because it was affected 
by errors induced by the location of the innermost boundary of the hydrodynamics grid 
($r^*=-3M$), as we have shown in Fig. \ref{label:fig5}. 
An indirect validation of the current estimation of energy
emission comes from an inspection of the findings of Shapiro and Wasserman \cite{shapiro82}.
These authors compute the total energy radiated in gravitational waves (while here we restrict 
to the $\l=2$ multipole) from non-spherical dust clouds  falling into a black hole from infinity. 
For any of the models considered, they find that the energy released in GWs is always 
smaller by at least two orders of magnitude with respect to the point--particle limit, 
with thinner clouds more efficient than wider ones.

\subsection{Neutron star simulations}
\label{ResultsStar}

We turn now to consider the case of quadrupolar perfect fluid shells accreting onto neutron
stars. Two neutron star models are considered, models A and B, whose characteristics (mass 
and radius) have been described in Sec.~\ref{stellar_model} above. Initially,
the quadrupolar shell is surrounded by a background fluid (an ``atmosphere") satisfying the 
stationary and spherically symmetric Michel solution~\cite{michel}. The initial rest mass 
density profile is given by Eq.~(\ref{shell}), where $\rho_0$ is now the profile consistent 
with the Michel solution. As in the black hole case, the mass of the shell is 
$\mu=0.01M$, which corresponds to a maximum density of 
$\rho_{\mathrm{max}}\sim 3.5\times 10^{-6}$~$\mathrm{km}^{-2}$ when we fix its width to 
$\kappa=1$. The (inhomogeneous) density profile of the atmosphere $\rho_{0}$ 
is roughly three orders of magnitude lower than $\rho_{\mathrm{max}}$. The shell obeys a 
polytropic ($p={\cal K}\rho^{\gamma}$) EOS with \hbox{${\cal K}=0.01$ $\mathrm{km}^{2/3}$} 
and $\gamma=4/3$. The initial internal energy profile is obtained from $\rho$ and $p$ 
through the first law of thermodynamics as $e=p/[(\gamma-1)\rho]$. The shell is initially 
at rest at $r_0=20$ km, and the GW signal is extracted at $r_{\mathrm{obs}}=250$ km.

\begin{table}[t]
\caption{\label{label:table2}Total energy emitted in gravitational waves for shells accreting
onto a Schwarzschild black hole from different distances $r_0$.}
\begin{ruledtabular}
\begin{tabular}{cccc}
&$r_0/M$      & $(2M/\mu^2)E^{20}$  & \\
\hline
& 7.5      & $1.09  \times 10^{-4}$ & \\
& 15       & $9.49  \times 10^{-5}$ & \\
& 30       & $7.20  \times 10^{-5}$ & \\
\end{tabular}
\end{ruledtabular}
\end{table}

\begin{figure*}[t]
\begin{center}
\includegraphics[width=85 mm,height = 80 mm]{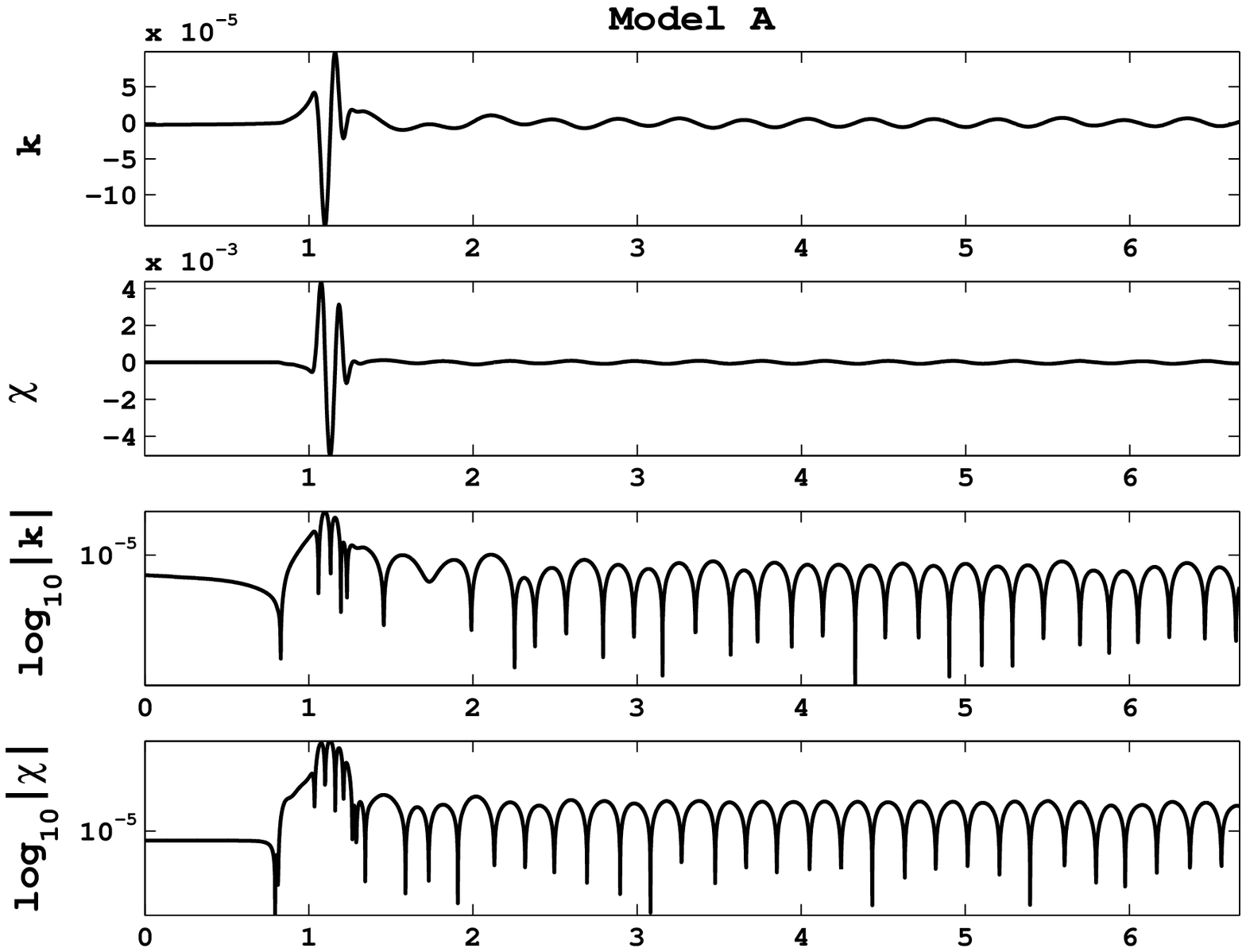}\qquad
\includegraphics[width=85 mm,height = 80 mm]{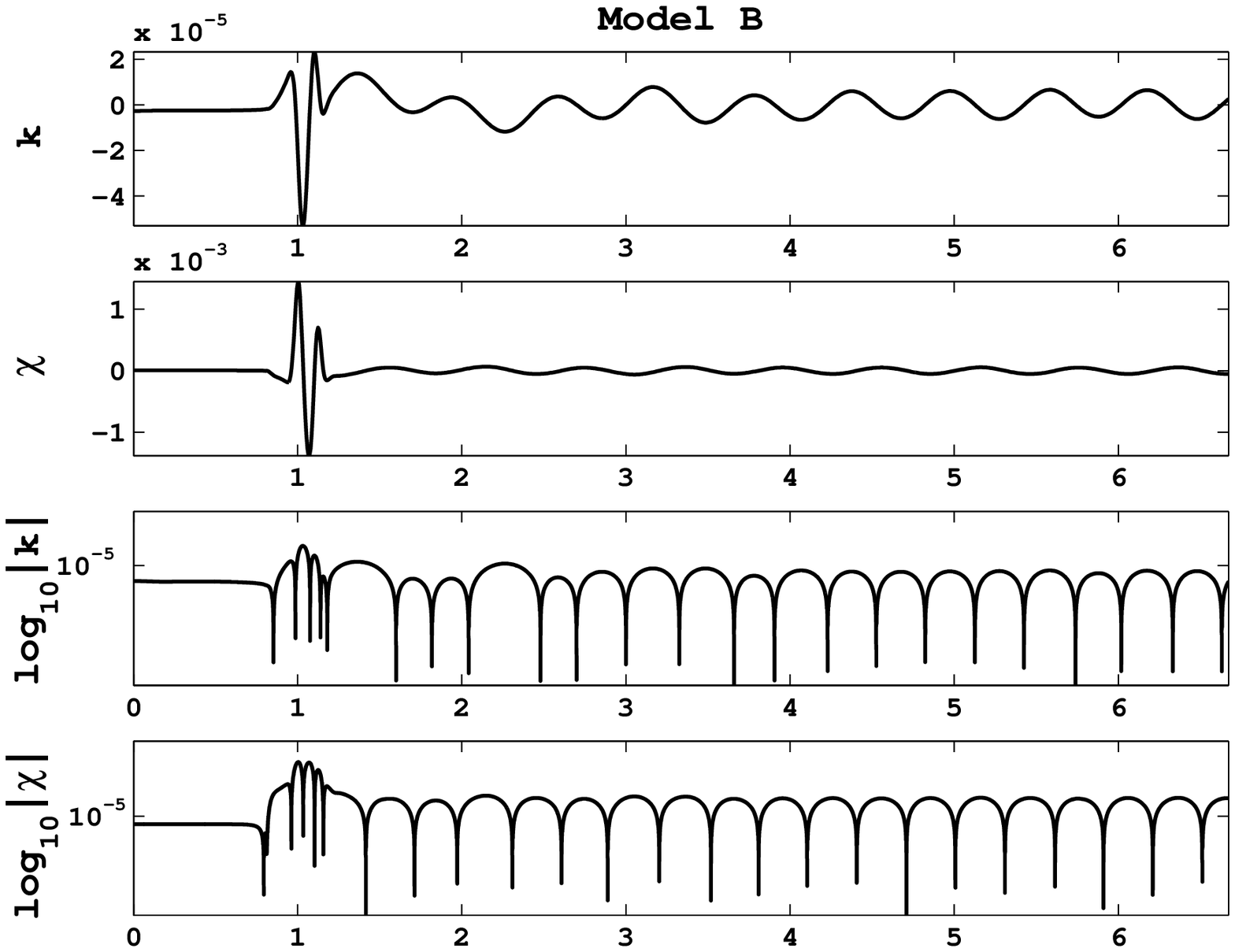}\\
\includegraphics[width=85 mm,height = 64 mm]{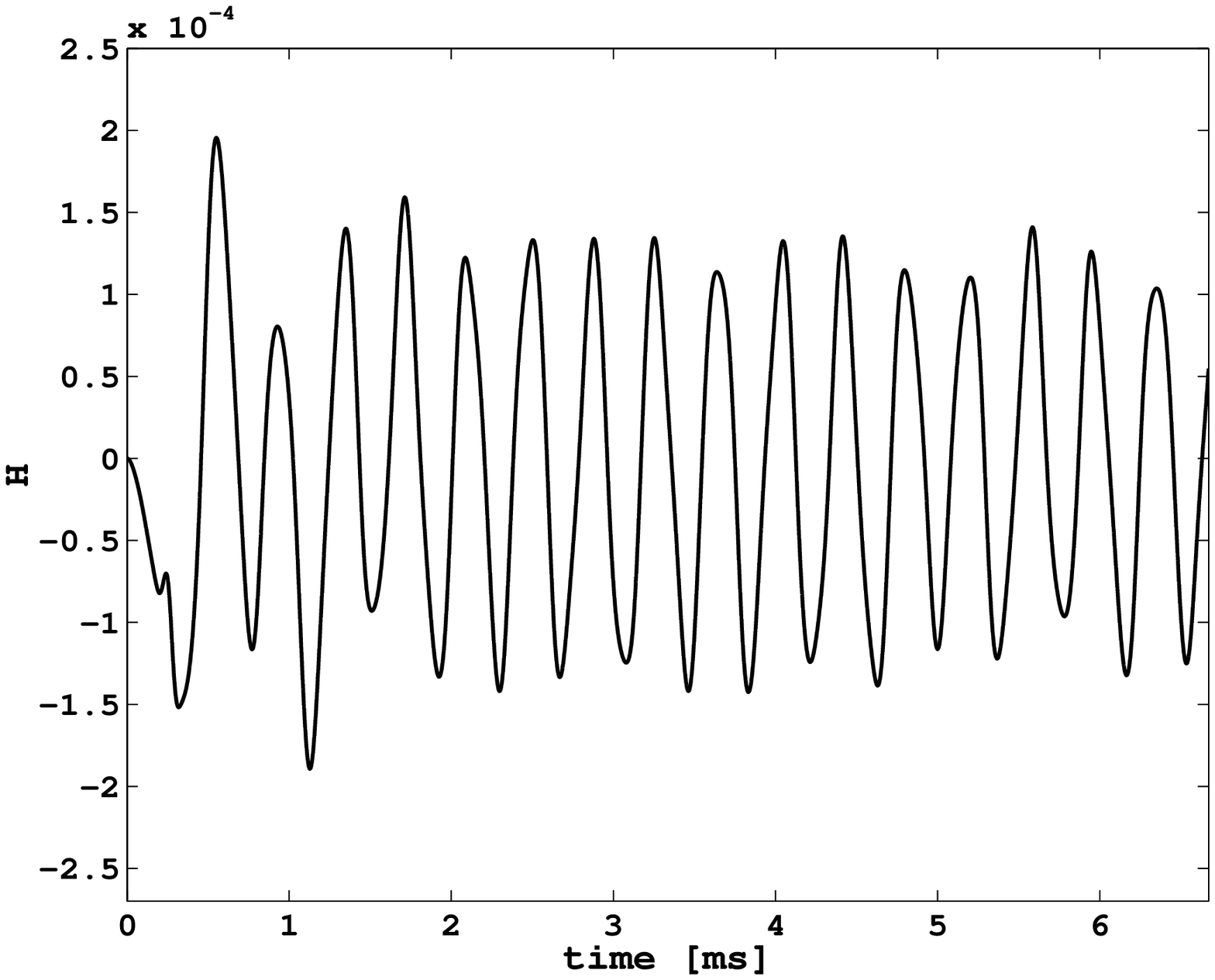}\qquad
\includegraphics[width=85 mm,height = 64 mm]{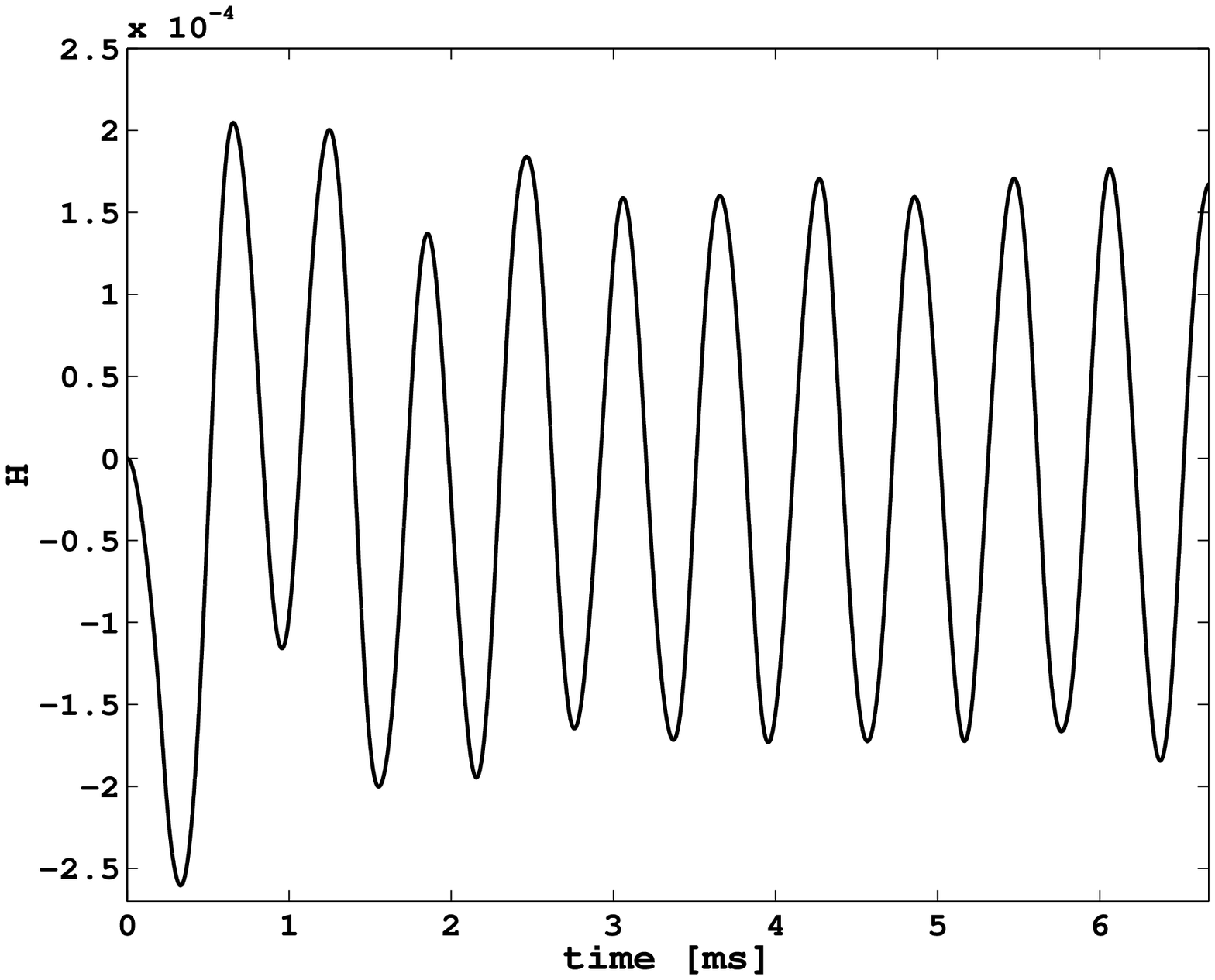}
\caption{\label{label:fig7}
Time evolution of the metric variables $k$ and $\chi$ and of the enthalpy perturbation $H$ for
the two neutron star models considered. The left panels correspond to model A and the right panels
to model B. The burst of gravitational radiation and the subsequent metric and fluid oscillations
are clearly identified for both stellar models. The infalling perfect fluid shell has an initial
compactness $\kappa=1$ and is located at a distance $r_0=20$ km. Note that, contrary to the previous
figures, the time is now given in ms.}
\end{center}
\end{figure*}
\begin{figure*}[t]
\begin{center}
\includegraphics[width=80mm,height=72 mm]{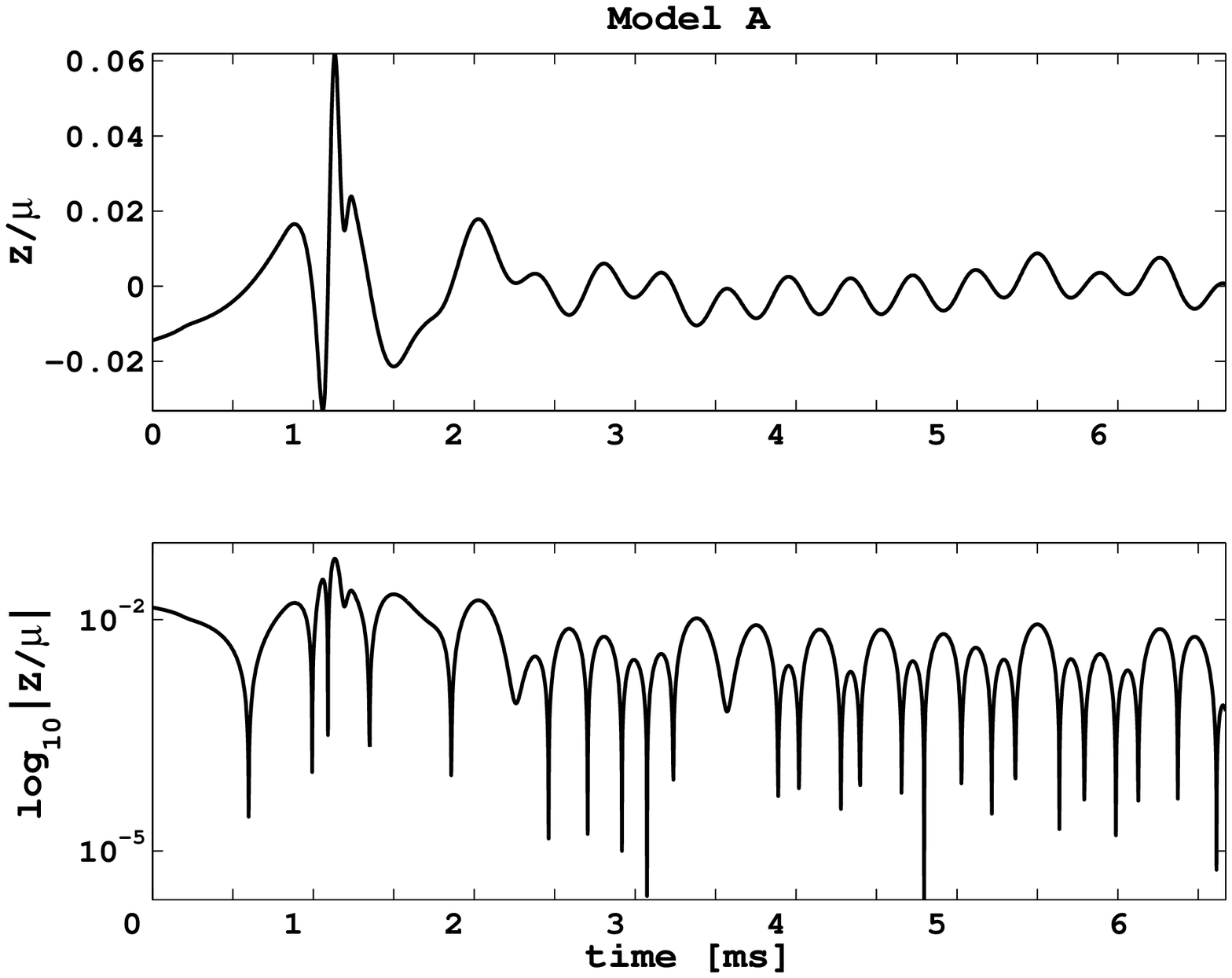}\qquad
\includegraphics[width=80mm,height=72 mm]{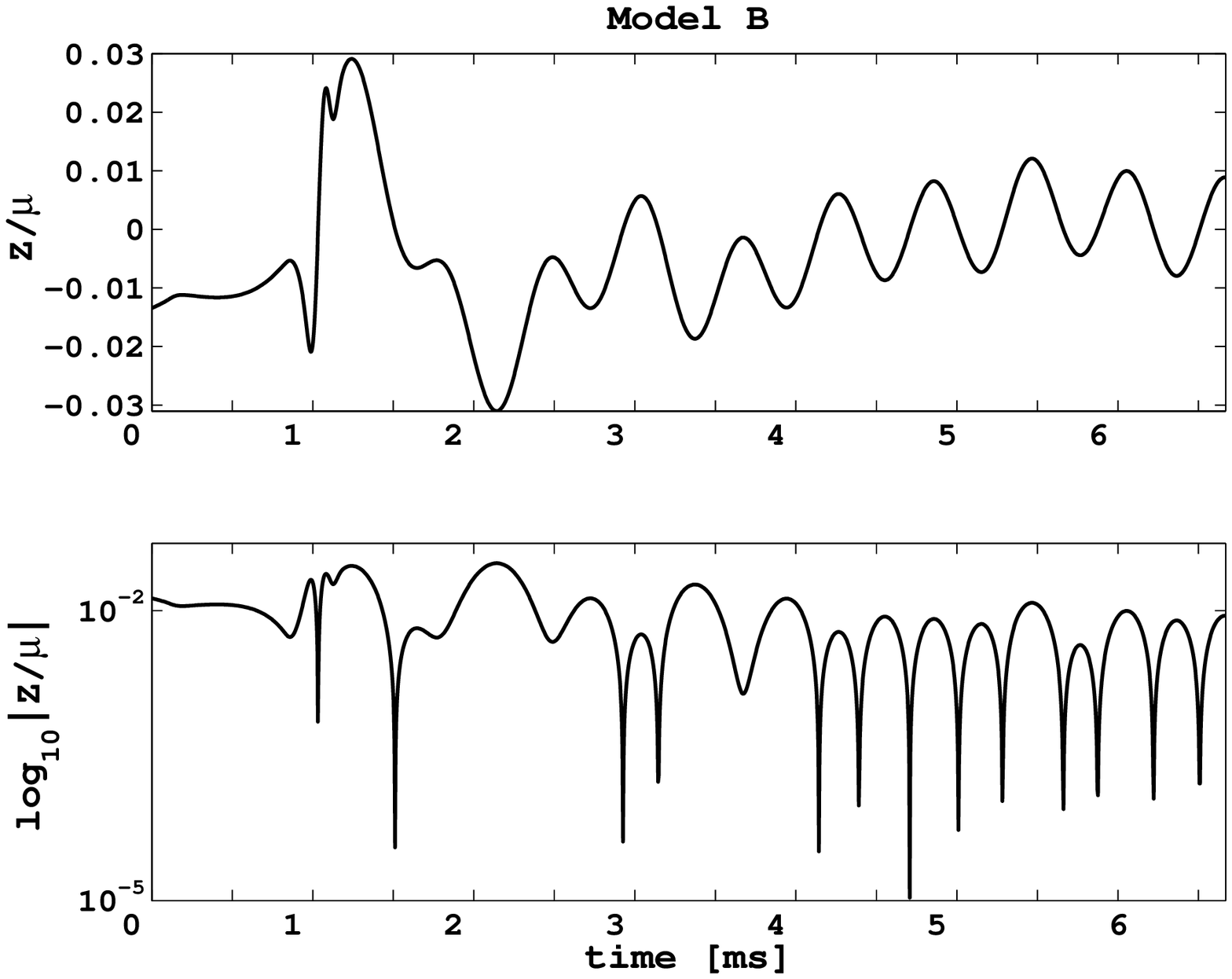}\\
\caption{\label{label:fig8}
Time evolution of the Zerilli--Moncrief function for the two stellar models considered. In 
close qualitative agreement with the results from gravitational core collapse simulations
\cite{Dimmelmeier}, the burst in the Zerilli--Moncrief function is associated with the moment 
of the bounce of the accreting shell at the stellar surface, and occurs at a retarded time 
of about $1$ ms.} 
\end{center}
\end{figure*}

We first present an overview of a typical evolution for both models, in order to have an 
immediate insight on the dependence of the gravitational wave signal on the compactness of 
the star. As a result of the reflecting boundary conditions imposed at the surface of the star and
the existence of an extended atmosphere, the accretion process is now followed by the formation 
of shock waves which propagate off the stellar surface. As in the case of the black hole, 
the impact of the shell perturbs the star and triggers its quasinormal modes of pulsation.

Figure~\ref{label:fig7} displays the time evolution of the metric functions 
$\chi$ and $k$, as well as the enthalpy perturbation $H$, for both models. Model A 
(B) is presented in the left (right) panels. In this figure 
we show the all the variables and the logarithms of the absolute values of the metric functions, 
in order to highlight the oscillating modes of the two stars. Correspondingly, Fig.~\ref{label:fig8} 
exhibits the time evolution of the Zerilli--Moncrief function for both models, in linear 
and logarithmic scales, computed at every time step according to Eq.~(\ref{DefZeta}). In 
the behavior of $Z$, as well as in the time evolution of $\chi$ and $k$ shown in 
Fig.~\ref{label:fig7}, the three phases discussed in the preceding section for the black 
hole case are also visible: first, the infalling phase, when the bulk of the shell is 
evolving outside the star, gradually approaching it, which is characterized by a steady 
increase of the amplitude of the signal. This phase is very short as the shell is 
initially located very close to the stellar surface. Second, a burst--like peak appears 
in the GW signal, which, as found in relativistic simulations of gravitational core 
collapse~\cite{Dimmelmeier}, coincides with the moment when the shell reaches the surface, 
creating a shock wave which propagates off the surface. Finally, there is the ringdown phase, 
characterized by a GW signal which is not exactly monochromatic as a result of the 
complex interaction between the gravitational field of the star and the layers of fluid 
captured on top of the stellar surface in the process of readjusting themselves to a new 
stationary solution. The duration of the ringdown phase is now much longer than for the 
Schwarzschild black hole case discussed previously, as the damping time of the fundamental 
mode of the fluid is considerably larger. We note that despite the waveforms obtained in 
our simulations showing a remarkable resemblance with those obtained in core collapse 
simulations~\cite{Dimmelmeier}, there are also important differences in the post--bounce 
phase dynamics, as the ringdown of the neutron star lasts for much longer times. In our 
idealized setup, with a perfect fluid, these pulsations are not quickly damped by the 
existence of a dense envelope surrounding the star, as happens in the core collapse situation.

The bulk of the accreting matter (the center of the shell) reaches the stellar surface 
at \hbox{$t_A\sim 0.2$} ms for model A and \hbox{$t_B\sim 0.14$} ms for model B. At 
$r_{\mathrm{obs}}$, where the observer is located, 
the GW signals generated by these events are delayed in time by 
\begin{align}
\Delta t=\bigg[r_{\rm obs}-R+2M\log\left(\dfrac{r_{\rm obs}-2M}{R-2M}\right)\bigg]\;.
\end{align}
Therefore, the signal generated by the matter bouncing back at the stellar surface
reaches the observer at times $t_A^{\rm obs}=t_A+\Delta t_A\approx 1.06\,{\rm ms}$
and $t_B^{\rm obs}=t_B+\Delta t_B\approx 0.97\,{\rm ms}$. These values are consistent 
with the results of Fig.~\ref{label:fig7}, where the waveforms of $k$ and $\chi$ are found 
to show bursts of large amplitude followed by highly damped oscillations sometime around these 
values. This observation is particularly confirmed in the logarithmic plots of the GW signals. 
After the short-lived ringing phase lasting for a fraction of half a millisecond after the burst, 
only the fundamental oscillation mode of the star is visible in all variables plotted in 
Figs.~\ref{label:fig7} and~\ref{label:fig8}.

\begin{figure*}[t]
\begin{center}
\includegraphics[width=80 mm,height=72 mm]{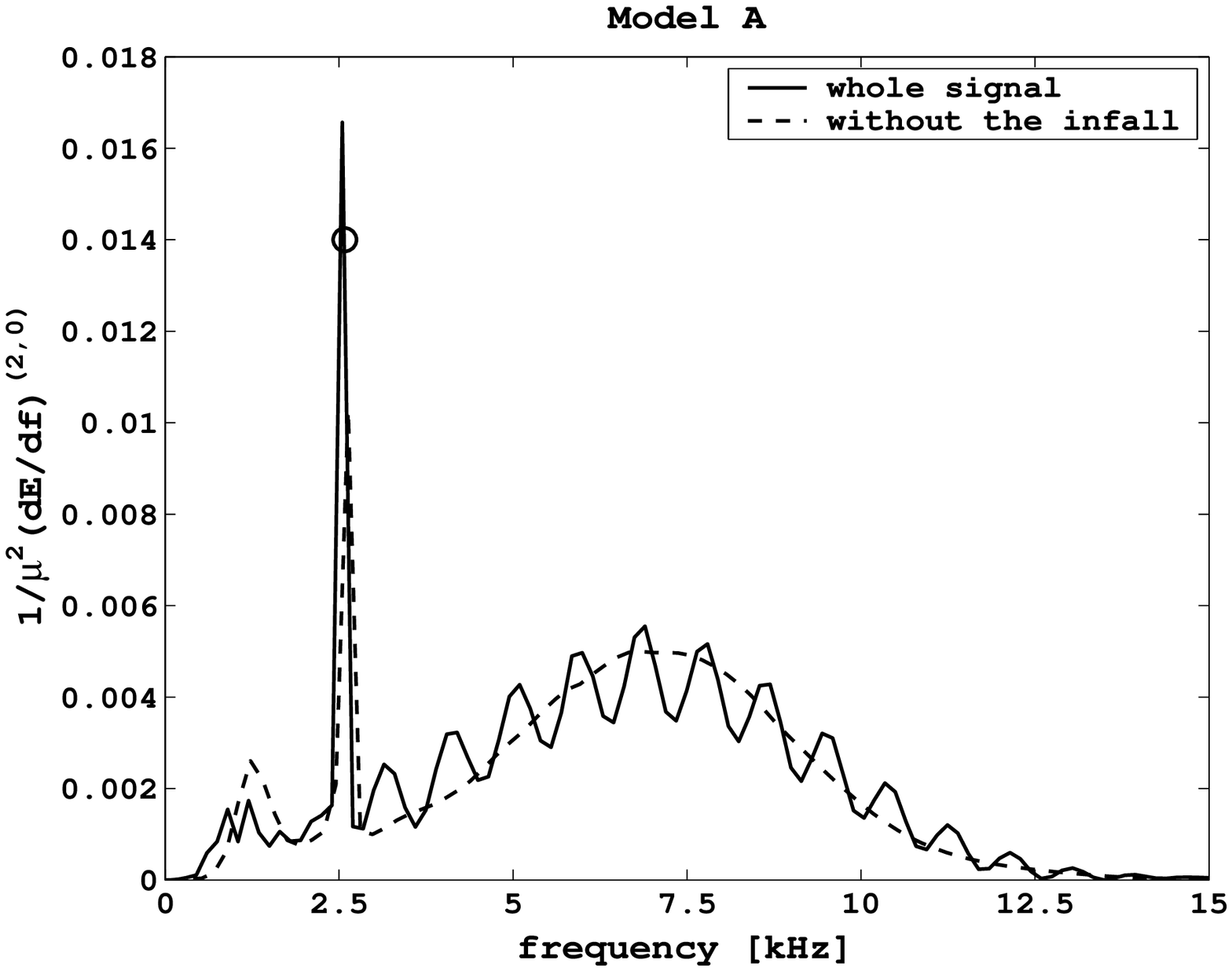}\qquad
\includegraphics[width=80 mm,height=72 mm]{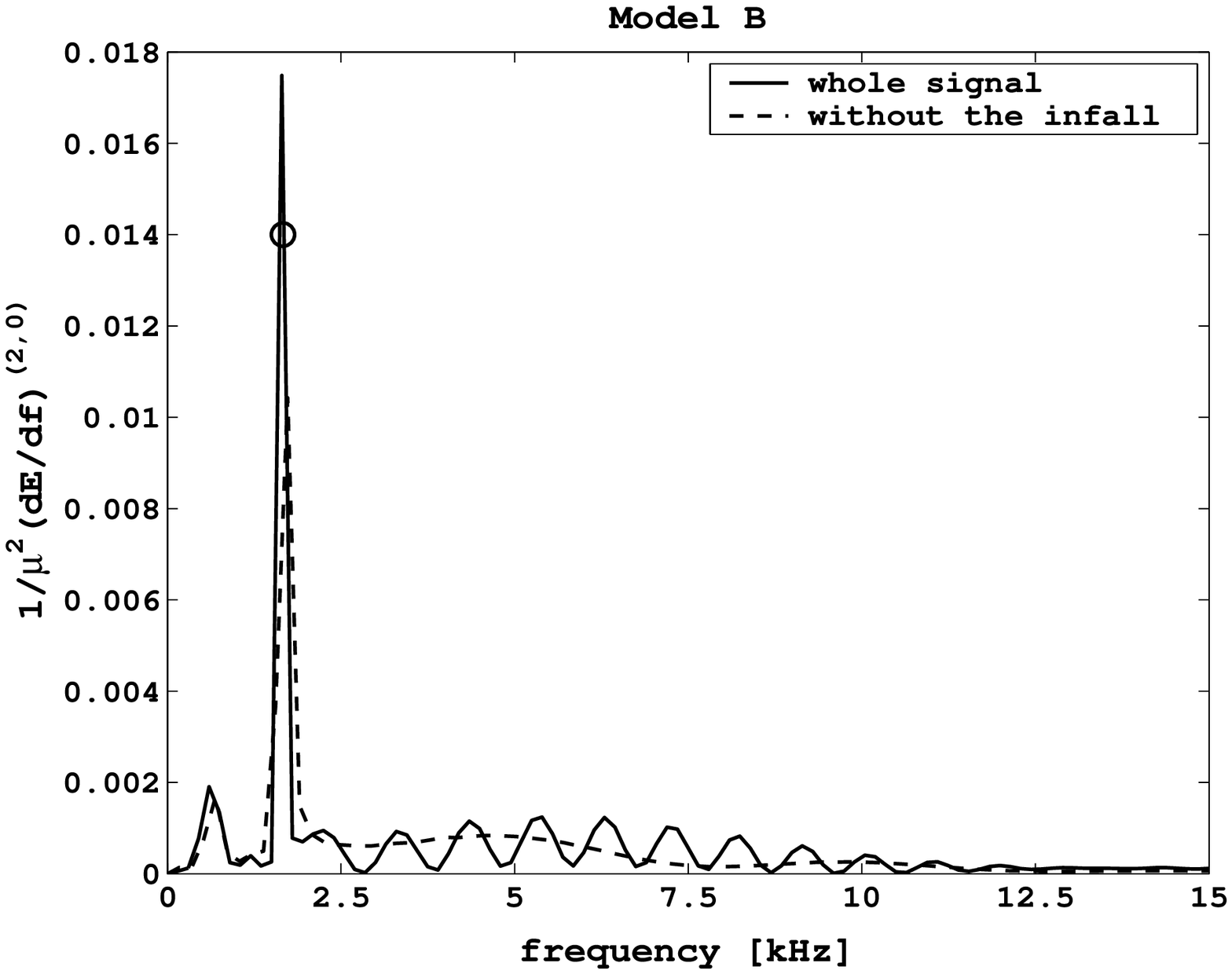}
\caption{\label{label:fig9}
Energy spectra corresponding to the time evolution of the Zerilli--Moncrief function 
depicted in Fig.~\ref{label:fig8}. Model A (B) is shown on the left (right) panel. The
solid lines show the spectra for the entire signal while the dashed lines are obtained
removing that part of the waveform which corresponds to the infall phase of the shell.
The spectra of the entire signal show characteristic interference fringes, qualitatively
similar to those found in the black hole case, with a large amplitude peak standing at
the frequency of the $f$ mode. The value of the frequency of the $f$ mode from
Table~\ref{label:table1} is indicated by a circle. The spectra of the truncated signals
show no evidence of the interference patterns.}
\end{center}
\end{figure*}

The qualitative behavior found in the GW emission is the same for both models. Only 
quantitative differences appear in the amplitudes at the maximum, which are systematically 
larger by roughly a factor of 2 for the more compact model (A). This difference in amplitude 
becomes more apparent in the energy spectra of the Zerilli--Moncrief function shown in 
Fig.~\ref{label:fig9}. The solid lines in this figure are obtained by Fourier 
transforming the complete signal, i.e.~also including the contributions from the shell 
infall phase which precedes the burst. Correspondingly, the dashed lines are obtained 
from truncated waveforms, in which we only take into account the contribution from the 
beginning of the burst ($\sim 0.9$ ms) onward. We notice that, despite the short evolution 
times of our simulations, the $f$ mode frequency is very well identified in the spectra, 
the relative difference with respect to the values listed in Table~\ref{label:table1} 
being roughly of 1\%. The value of the $f$ mode frequency is indicated with a 
circle in Fig.~\ref{label:fig9}. Furthermore, it is worth stressing that we obtain 
the same qualitative spectra as for the black hole case -- a complex pattern with 
interference fringes with the addition, in the neutron star case, of a high peak standing 
at the frequency of the $f$ mode. As we discussed in the preceding section for a 
Schwarzschild black hole, the comparison between the spectra obtained from the total 
and the truncated Zerilli signal shows that the interference fringes are produced by 
the interaction of the infalling fluid and the star.

There are other small differences between the two stellar models. The more compact 
model (A) is more efficient at high frequencies than model B. This is directly correlated 
with the large amplitudes attained by the peaks of the Zerilli--Moncrief function in the time 
domain (cf.~Fig.~\ref{label:fig8}). The broadband spectrum of model A presents one broad 
peak with a maximum at about $7$ kHz, which is however too low to be identified with the 
first $w$ mode (see Table~\ref{label:table1}). As mentioned before, its origin should be related 
to the motion of the fluid shell and its interference with the gravitational field of the star, 
or, in other words, on the reflection of the GWs pulse associated with the shell distribution 
with the ``external" Zerilli potential. Thus, this broad feature depends on details of the 
accretion dynamics rather than on the intrinsic characteristics of the star.  

\begin{figure}[t]
\begin{center}
\includegraphics[width=80 mm,height=70 mm]{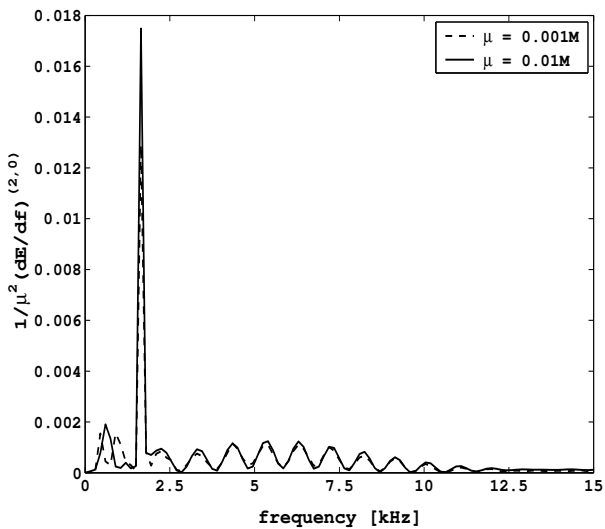}
\caption{
\label{label:fig10}
Energy spectra for two fluid shells differing on the initial mass $\mu$ and infalling from a 
distance of $r_0=20$ km onto stellar model B. The solid line corresponds to $\mu=0.01M$ and 
the dashed line to $\mu=0.001M$. The $f$ mode excitation and the high frequency part of the
spectra coincide to high precision for the two shells considered. Only the low frequency
peak is affected by the different mass of the shell.
}
\end{center}
\end{figure}

We note that in both spectra there appears a small amplitude peak at frequencies lower than
that of the $f$ mode for each model. This second peak is associated with oscillations of that
part of the external fluid that has been gravitationally captured by the central neutron star 
as a result of accretion. In order to illustrate this affirmation, we plot in Fig.~\ref{label:fig10} 
the energy spectra corresponding to two fluid shells which only differ on the mass ($\mu=0.01M$ 
and $0.001M$), falling onto stellar model B from the same distance, $r_0=20$ km. The $f$ mode 
is also properly excited for the less massive shell. Furthermore, the interference fringes in 
the high frequency part of the spectra coincide to high precision for the two shells considered, 
after normalizing to the corresponding shell masses. However, the structure of the low frequency 
peak is the only feature of the spectra which is influenced by the mass of the shell. 
We argue that the existence of this unphysical low frequency peak is an artifact produced by the 
boundary conditions. Notice, once more time, that we model the surface of the star as a hard surface, 
with reflecting boundary conditions. In a realistic scenario, the accreted matter would not simply 
bounce at the stellar surface, but it would rather interact with the neutron star envelope, resulting 
in heating and suffering nuclear reactions, until it is reabsorbed by the star.

As we did for the black hole case, let us now close this section by computing the total energy 
emitted in gravitational waves for the two stellar models considered. For doing so we integrate 
in frequency the spectra appearing in Fig.~\ref{label:fig9}. The result of the integration yields
\begin{align}
E_{A}^{20}&\simeq  
3.02\times 10^{-8}\;M_{\odot}c^2\;,\\
E_{B}^{20}&\simeq
8.36\times 10^{-9}\;M_{\odot}c^2\;.
\end{align}
Model A is hence slightly more efficient than model B concerning GW emission. We note in 
passing that these values are as small as those found in core collapse simulations~\cite{Dimmelmeier}, 
although such a comparison only makes sense in qualitative terms. Finally, the mass of the shell 
radiated in gravitational waves is
\begin{align}
E_{A}^{20}&\simeq 2.15\times 10^{-6}\;\mu\;,\\
E_{B}^{20}&\simeq 5.97\times 10^{-7}\;\mu\;.
\end{align}

\section{Conclusions}
\label{conclusions}

In this paper we have presented a detailed analysis of the gravitational radiation 
induced by the anisotropic accretion of quadrupolar (dust or perfect fluid) shells onto 
non--rotating black holes and neutron stars. The numerical framework for our simulations is 
based upon a {\it hybrid} procedure in which the linearized equations describing metric and 
fluid perturbations are coupled to a fully nonlinear hydrodynamics code that calculates the 
motion of the accreting matter. These equations are integrated numerically in axisymmetry 
using advanced computational techniques. The numerical schemes developed have proved to be 
stable and highly accurate. Regarding the perturbation equations, the two main technical 
changes with respect to previous works reported in the literature are the use of the 
Zerilli-Moncrief function and the in-built conservation of the Hamiltonian constraint. 
The hydrodynamics equations are solved using high-resolution shock-capturing schemes based 
upon approximate Riemann solvers. We have shown that a perturbative approach can be used as 
a very effective tool to understand the basic gravitational physics operating in interesting 
astrophysical situations, extending the information which can be gained from the study of 
point-like particles infalling onto black holes or orbiting around them. In this context, 
our hybrid approach can be extremely useful to understand the gravitational radiation from 
astrophysical systems, complementing the whole machinery of full numerical relativity.

The simple fluid configurations analyzed in this paper ascertain, however, that the effects 
of the extended structure of the accreting matter are indeed highly relevant for  
gravitational wave emission. In the black hole case, most of the energy is released at 
frequencies lower than that of the fundamental QNM of the black hole, the spectrum consisting
of a complex pattern, mostly produced during the accretion process rather than in the 
ringdown phase. Therefore, the GW emission that could be expected from a realistic astrophysical
scenario in which accretion operates (e.g.~after gravitational collapse of a massive star) 
is more likely to be a sort of interference fringe, covering low frequencies, than a single 
monochromatic peak at the frequency of the fundamental mode of the black hole. This result, 
which went unnoticed in earlier hydrodynamical simulations by~\cite{PapadopoulosFont}, was 
already observed by Lousto and Price~\cite{LoustoPrice} in the case of a point--like particle 
falling onto a black hole (see Appendix~\ref{appendixB}), a situation amenable to semi--analytic 
investigation. However, we have shown that the appearance of interference fringes is very 
much amplified when the accreting fluid is an extended shell of finite size which, in turn,
reduces the amount of energy which is released in gravitational waves to some two orders of
magnitude below the point--particle value. It is interesting to notice that ground--based 
interferometric detectors attain the maximum sensitivity at frequencies considerably lower 
than the QNMs of stellar mass black holes. For this reason they are usually not considered 
as optimal sources for detection. However, to the light of our findings, the process of 
accretion appears to be quite more effective at frequencies about a factor of 2--3 lower than 
that of the black hole fundamental mode, and therefore the chances of detecting gravitational 
wave signals from such scenarios may be larger than expected. It must be stressed that the 
interference pattern does not happen only because of the GW content of the initial data, 
but it is a distinctive feature due to the extended size of the object, resulting from the 
interaction between the infalling matter and the backscattered waves.

In the neutron star case the qualitative results are similar, but a considerable part of the 
energy is emitted at the frequency of the fundamental mode. We have shown that the $f$-mode 
of the star is the only one excited at significant levels, and that the high frequency spectrum
is quite sensitive to the spatial distribution of the accreting matter, making the contribution
of the spacetime $w$-modes of the star difficult to be identified. The waveforms obtained show 
a remarkable resemblance with those obtained in core collapse simulations~\cite{Dimmelmeier}, 
despite the fact that we are considering a very different scenario. The main difference is that 
in our case the ringdown of the neutron star lasts for much longer times: the pulsations are not 
quickly damped by the existence of a dense envelope surrounding the star, as happens in the core 
collapse situation.

The results reported in this paper can partly be considered as a necessary assessment of our 
numerical approach in anticipation of the study of more interesting astrophysical scenarios, 
namely the excitation of QNMs from perfect fluid thick accretion tori orbiting around compact 
objects (see e.g.~\cite{zanotti,daigne} and references therein), which will be presented 
elsewhere~\cite{diazetal}.

\section*{ACKNOWLEDGMENTS}
We thank L.~Villain, E.~Berti, V. Ferrari, R.~De Pietri, and E. Onofri for discussions, 
suggestions and assistance, and K.~D.~Kokkotas and L.~Rezzolla for a critical reading 
of the manuscript. This work has been supported by the EU Programme ``Improving the 
Human Research Potential and the Socio-Economic Knowledge Base'' (Research Training 
Network Contract HPRN-CT-2000-00137) and the Spanish MCyT grant AYA 2001-3490-C02-01. 
J.A.P.~is supported by a Ram\'on y Cajal contract from the Spanish MCyT. All the 
computations were performed on the INFN Albert100 cluster for numerical relativity 
of the University of Parma.

\appendix

\section{General source term for the Zerilli-Moncrief equation}
\label{appendixA}

In this appendix we derive the source term given in Eq.~(\ref{ZerilliSource}) for a 
general stress--energy tensor. We use the normalization of the Zerilli--Moncrief 
function $\phi$ given in Ref.~\cite{GundlachII}, although the function we evolve 
in the numerical simulations is rescaled as $Z=2 \phi /\lambda$. 
The inhomogeneous equation written using the frame derivatives notation 
of \cite{GundlachII} reads 
\begin{align}\label{PhiEq}
-\ddot{\phi}+\phi''+\nu \phi'-V_{\phi}\phi=S_\phi\;,
\end{align}
where $V_{\phi}$ is the potential and $S_{\phi}$ the source term. 
As shown by Moncrief \cite{Moncrief}, 
$\phi$ can be written in terms of $k$ and $\chi$ as 
\begin{align}\label{DefPhi}
\phi=A(r)\chi+B(r)k+C(r)k'\;,
\end{align}
where 
\begin{align}
A &= \dfrac{2r^2 e^{-2b}}{(\lambda-2)r+6M}\label{Ac}\;,\\
B&=\dfrac{r(r\lambda+2M)}{(\lambda-2)r+6M}\label{Bc}\;,\\
C&=-rAe^b\label{Cc}~.
\end{align}
The derivation of the source term $S_\phi$ follows from the knowledge of the  
source terms in the evolution equations for $\chi$ and $k$ that can be written 
in vacuum~\cite{GundlachII}. Using Eq.~(\ref{DefPhi}) in Eq.~(\ref{PhiEq}) we get 
\begin{align}
-\ddot{\phi}+\phi''+\nu \phi'&=A(-\ddot{\chi}+\chi'')-B\ddot{k}\nonumber\\
&+\left[B+2C'+\nu C\right]k''\nonumber\\
&+C\left[-(k')\ddot{\;}+k'''\right]\nonumber\\
&+L\left(k,\chi,k',\chi'\right)\;,
\label{phieq}
\end{align}
where $L$ is a linear operator acting on $k$, $\chi$, $k'$ and $\chi'$, whose explicit 
form is not relevant for the computation of the source. In fact, in Ref.~\cite{GundlachII} 
it is shown that all terms which are {\it linear} in the fields $\chi$ and $k$  and 
their first--order spatial frame derivatives merge together to build the Zerilli potential. 
The frame derivatives do not commute \cite{GundlachI},
\begin{align}
(\dot{k})'-(k')\dot{}=-\nu\dot{f}\;,
\end{align}
so that 
\begin{align}
(k')\ddot{\;}=(\ddot{k})'+2\nu\ddot{k}\;,
\end{align}
$\dot{\nu}=0$ being on a static background. Equation (\ref{phieq}) becomes then
\begin{align}
-\ddot{\phi}+\phi''+\nu \phi'&=A\left(-\ddot{\chi}+\chi''\right)+B\left(-\ddot{k}+k''\right)\nonumber\\
&+\left(\nu C+2C'\right)k''-2\nu C\ddot{k}\nonumber\\
&+C\left(-\ddot{k}+k''\right)'+L\left(k,\chi,k',\chi'\right)\;.
\end{align}
%
\begin{figure*}[t]
\begin{center}
\includegraphics[width=80 mm,height=72 mm]{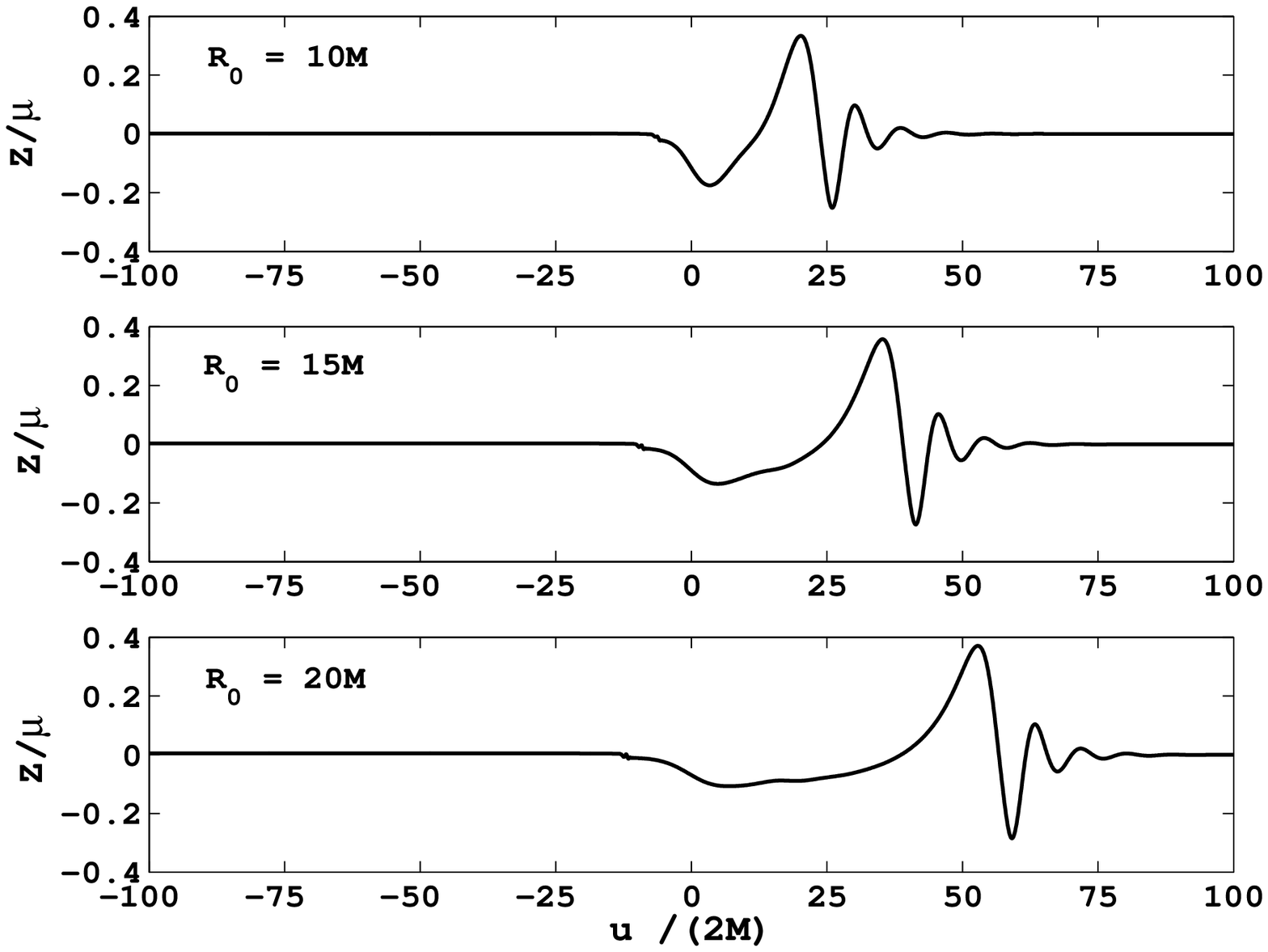}\qquad\;
\includegraphics[width=80 mm,height=72 mm]{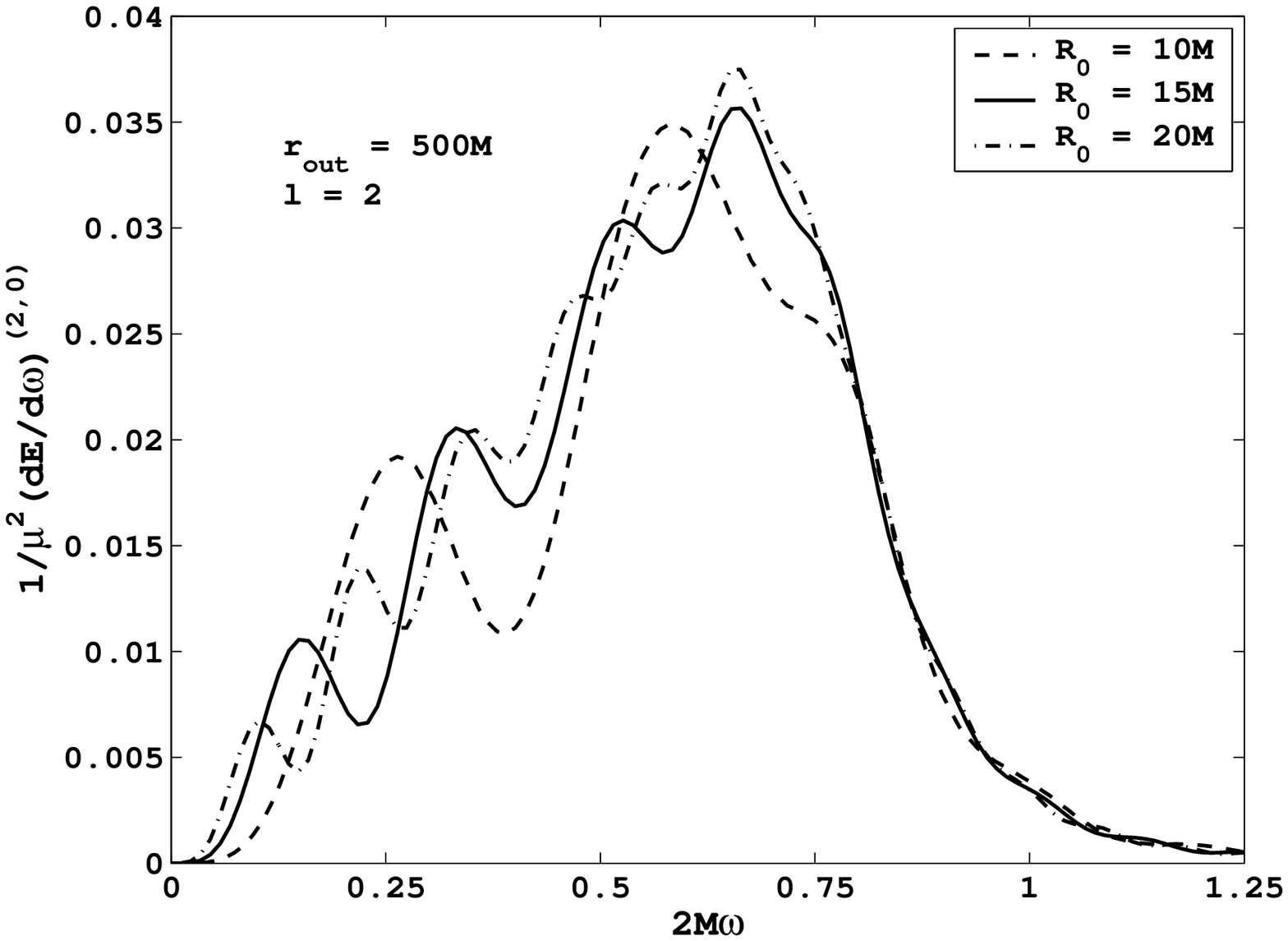}\\
\caption{\label{label:fig11}
Gravitational waveforms and energy spectra for a point particle falling radially onto a
Schwarzschild black hole from different distances. The left panels show the waveforms and
the right panel the corresponding spectral energy distributions. Excellent agreement is
found with the results of Lousto and Price \cite{LoustoPrice}. See text for further details.}
\end{center}
\end{figure*}
%
Next, we can write the evolution equations for $k$ and $\chi$ in a Schwarzschild spacetime
and the Hamiltonian constraint as 
\begin{align}
-\ddot{\chi}+\chi''&=L\left\{k,\chi,\chi'\right\}+T_{\chi}\\
-\ddot{k}+k''&=L\left\{\chi,k\right\}-\left(2W+\nu\right)k'+T_k\\
k'' &=L\left\{\chi,k,\chi',k'\right\}+T_{\cal H}\;,
\end{align}
with $T_{\chi}$, $T_k$ and $T_{\cal H}$ being the sources induced by
the matter distribution. From these equations one obtains
\begin{align}
\left(-\ddot{k}+k''\right)'=L\left\{k,\chi,k',\chi'\right\}-\left(2W+\nu\right)T_{\cal H}+T_k'\;,
\end{align}
so that the source of the Zerilli equation is found to be
\begin{align}\label{source}
S_{\phi}&=AT_{\chi}+(B+2\nu C)T_k\nonumber\\
&+\left[2C'-2(W+\nu)C\right]T_{\cal H}+CT_k'\;.
\end{align}
In polar radial coordinates the sources are, explicitly,
\begin{align}
T_{\cal H}&=-8\pi S_{\cal H}\label{tacca}\;,\\ 
T_{\chi}&=-16\pi e^{2a}S_{\chi}\label{tchi}\;,\\
T_{k}&=8\pi\bigg\{-e^{-2a}T_{00}^{\lm}+e^{-2b}T_{11}^{\lm}\nonumber\\
&\qquad\quad+4\dfrac{e^{-2b}}{r}T_1^{\lm}
-2\dfrac{e^{-2b}}{r}\left(T_{2}^{\lm}\right)_{,r}\nonumber\\
&\qquad\quad-\left(\dfrac{\lambda+2}{r^2}
-\dfrac{6e^{-2b}}{r^2}\right)T_2^{\lm}\bigg\}\label{tkappa}\;,
\end{align}
where $S_{\cal H}$ and $S_{\chi}$ have been defined in Sec.~\ref{StarPerturbations}.
When Eqs.~(\ref{tacca})--(\ref{tkappa}) and  the definitions of the coefficients 
(\ref{Ac})--(\ref{Cc}) are replaced in Eq.~(\ref{source}), we get 
$S_\phi=\lambda S_z/2$, with $S_z$ given by Eq.~(\ref{ZerilliSource}).

The projections of the source stress-energy tensor onto the basis of 
the spherical harmonics is accomplished as follows. From the orthogonality
properties of the harmonics,
\begin{align}
&\int d\Omega Y^*_{\lm}Y_{\l'm'}=\delta_{\l\l'}\delta_{mm'}\;,\\
&\int d\Omega Y^*_{\lm,a}Y_{\l'm',b}\gamma^{ab}=\lambda\delta_{\l\l'}\delta_{m m'}\;,\\
&\int d\Omega Z_{ab}^{*\l'm'}Z^{ab}_{\lm}=\frac{\lambda\left(\lambda -2\right)}{2}\,
\delta_{\l\l'}\delta_{m m'}\;.
\end{align} 
we obtain the coefficients of the expansion of $t_{\mu\nu}$,
\begin{align}
T_{AB}^{\lm}&=\int d\Omega\,t_{\mu\nu}Y^*_{\lm}\;, \qquad A,B,\mu,\nu=0,1,\\
T_0^{\lm}&=\dfrac{1}{\lambda}\int d\Omega \left[t_{02}Y^*_{\lm,\vartheta}
-\dfrac{{\rm i}m}{\sin^2\vartheta}\,t_{03}Y^*_{\lm}\right]\;,\\
T_{1}^{\lm}&=\dfrac{1}{\lambda}\int d\Omega \left[t_{12}Y^*_{\lm,\vartheta}
-\dfrac{{\rm i}m}{\sin^2\vartheta}\,t_{13}Y^*_{\lm}\right]\;,\\
T_{3}^{\lm}&=\dfrac{1}{2r^2}\int d\Omega \left[t_{22}+\dfrac{1}{\sin^2\vartheta}t_{33}\right]Y^*_{\lm}\;,
\end{align}
and
\begin{align}
T_{2}^{\lm}&=\dfrac{2}{\lambda(\lambda-2)}
\int d\Omega\,\dfrac{t_{22}}{2}\,W^*_{\lm}+\dfrac{2 t_{23}}{\sin\vartheta}\,X^{*}_{\lm}\\
&+\dfrac{t_{33}}{\sin^2\vartheta}\left[-m^2Y^*_{\lm}+\dfrac{\lambda}{2}\sin^2\vartheta Y^*_{\lm}
+\frac{1}{2}\sin2\vartheta Y^*_{\lm,\vartheta}\right],\nonumber
\end{align}
where, following Regge and Wheeler \cite{ReggeWheeler}, we have defined the functions
\begin{align}
X^*_{\lm}&=2{\rm i}m\left[\cot\vartheta Y^*_{\lm}-Y^*_{\lm,\vartheta}\right]\;,\\
W^*_{\lm}&=2Y^*_{\lm,\vartheta\vartheta}+\lambda Y_{\lm}^*\;.
\end{align}
In the general case where $t_{\mu\nu}$ describes a complex source corresponding to a
general distribution of matter evolving dynamically, these integrals have to be evaluated 
numerically. The analytic computation can be done only in some simple cases, such as the
particular case when the source is a test-mass body moving along a geodesic of Schwarzschild 
spacetime~\cite{RuoffII}.

\section{Point--like particles radially falling onto black holes.}
\label{appendixB}

In this appendix we reexamine the simplified scenario of a point particle radially falling 
onto a Schwarzschild black hole. This is done with two purposes: first, to test our 
perturbative numerical code with previous works and, second, to analyze the
similarities and differences with the case of accretion of extended fluid shells onto
black holes. The emission generated by infalling particles has been extensively studied 
in the past. The seminal calculation of the GW emission when a test particle falls from 
infinity \cite{DRPP} was later extended to non-radial trajectories by Detweiler and Szedenits 
\cite{DetweilerSzedenits}. In both cases the analysis was done in the frequency domain. 
A frequency component treatment based on Laplace transforms was also employed by Lousto 
and Price \cite{LoustoPrice} in the study of the emission from particles falling from 
finite distances. On the other hand, a treatment of the same problem in the time domain 
has just recently been approached by Martel and Poisson \cite{MartelPoisson}.

\begin{figure}[t]
\begin{center}
\includegraphics[width=80 mm,height=70 mm]{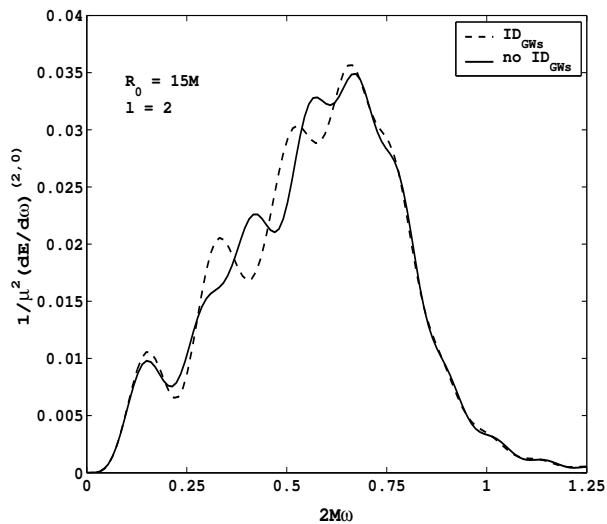}
\caption{\label{label:fig12} 
Effects of the gravitational wave content of the initial data on the energy spectra of a
point--like particle falling radially onto a black hole from $R_0=15M$. By eliminating the
initial contribution of GWs the amplitude of the bumps is strongly reduced. Qualitatively,
however, the effect is always present.}
\end{center}
\end{figure}

In order to test our numerical code we should be able to
reproduce those results reported in Refs.~\cite{LoustoPrice,MartelPoisson} within the current 
time domain approach.  A way to deal with a $\delta$-like source and with a discontinuous 
Zerilli-Moncrief function at the location of the particle was developed by 
Lousto and Price \cite{LoustoPriceII}, and later successfully applied in Ref.~\cite{MartelPoisson} 
following a time domain approach. However, we found it convenient and accurate enough for
our purposes to represent the $\delta$ function in the particle source terms by 
a narrow Gaussian written as
\begin{align}
\delta(r-R_0)\approx \dfrac{1}{\sigma\sqrt{2\pi}}\,e^{-(r-R_0)^2/2\sigma^2}\;,
\end{align}
with $\sigma\ll 1$. For a given resolution, we fix $\sigma=2\Delta r^*$, so that the smaller 
$\Delta r^*$, the better the approximation of the $\delta$ function is. The Zerilli-Moncrief 
equation is solved on an evenly spaced grid using a standard three-level leapfrog scheme.
We consider particles initially at rest falling from a finite distance $R_0$ to compare
with the results of Ref.~\cite{LoustoPrice}. Since the particle is falling along the \hbox{$z$ axis},
the system is axisymmetric and the only non-vanishing contribution for any $\l$ is the $m=0$ one. 
The source terms are specified accordingly and we consider explicitly just $\l=2$. 

As a test of our numerical method, we consider the particle falling from $R_0=10M$, $15M$, 
and $20M$, with the same initial setup of Ref.~\cite{LoustoPrice} (i.e.~including some initial 
GW content) and compute the evolution of the Zerilli-Moncrief function and its corresponding 
energy spectrum. Figure~\ref{label:fig11} shows the results of these simulations. The left 
panels show the temporal evolution of $Z$ normalized to the particle mass, while the right 
panel exhibits the corresponding energy spectra. The waveforms show a good agreement in 
amplitude and shape with those of Ref.~\cite{LoustoPrice}. The energy spectra for $R_0=10M$, $15M$, 
and $20M$ must be compared with Figs.~6(b), 4(b), and 4(c) of Ref.~\cite{LoustoPrice}, respectively.
The spectra show the same localized bumps due to the interference between the initial GW pulse 
and the GWs emitted by the particle during its motion. We notice quite good agreement for 
$R_0=10M$ and $R_0=15M$, while some small differences are found for $R_0=20M$.

Next, we analyze the contribution of the initial data on the power spectra. We have studied 
how the spectrum changes when the initial GW contribution is eliminated from the evolution. This 
is accomplished in the same way we used for the extended shells; that is, the particle is 
frozen at its location until the initial pulse has gone from the numerical domain, after 
which the evolution starts. Figure~\ref{label:fig12} compares the energy spectra emitted by 
a particle falling from $R_0=15M$ with (dashed line) and without (solid line) the initial 
GW contribution. The signal is extracted at $r=500M$ in the two cases. As suggested in a more 
general scenario in Ref.~\cite{MartelPoisson}, our results confirm that in the test particle 
case the bumps in the spectrum are mainly due to the spurious contribution of the radiation 
in the initial data. In fact, when removing the initial GW content, the amplitude of the bumps 
is reduced. Some modulation is, nevertheless, still present and the spectrum does not fully 
correspond to that of a pure QNM ringdown signal. Martel and Poisson~\cite{MartelPoisson} argued 
that this is to be interpreted as an interference effect as well, but between the waves emitted 
by the particle during its motion and those previously emitted and backscattered by the potential. 


\end{document}